%% file: main.tex
\documentclass{article}
\usepackage{authblk}
\usepackage[utf8]{inputenc}
\usepackage[a4paper, left=1in, right=1in, top=1in, bottom=1in]{geometry}
 
\usepackage{float}
\usepackage{graphicx} 
\graphicspath{{figure/}} 
\usepackage{abstract}
\usepackage{subcaption}
\usepackage{appendix} 
\usepackage{comment}
\usepackage{booktabs}
\usepackage{longtable}
\usepackage{amssymb}
\usepackage{amsmath,amsthm,mathtools}
\usepackage{hyperref}
\usepackage[backend=biber,style=numeric,sorting=none]{biblatex}
\usepackage{algorithm}
\usepackage{algpseudocode}
\usepackage{xcolor}
\usepackage{microtype}   

\theoremstyle{plain}

\theoremstyle{definition}

\theoremstyle{remark}

\usepackage{cleveref}
\Crefname{equation}{Eq.}{Eqs.}
\Crefname{figure}{Fig.}{Figs.}
\Crefname{table}{Table}{Tables}
\Crefname{algorithm}{Algorithm}{Algorithms}
\Crefname{section}{Sec.}{Secs.}
\Crefname{appendix}{Appendix}{Appendices}
\definecolor{commentgray}{gray}{0.45}
\algrenewcommand\algorithmiccomment[1]{%
  {\color{commentgray}\hfill$\triangleright$\ \textit{#1}}}
\newcommand{\mycomment}[1]{{\color{commentgray}\textit{#1}}}
\algrenewcommand\algorithmicindent{1.4em}
\newcommand{\Tx}{T_x}
\newcommand{\Ts}{T_s}
\newcommand{\GCC}{\mathcal{S}}
\newcommand{\Adeg}{\mathcal{A}}

\begin{document}

\title{The impact of the path ensemble on path percolation}

\author[1]{Yunhao Ding}
\author[1]{Andreas M\"{u}nch}
\author[1]{Renaud Lambiotte}
\affil[1]{Mathematical Institute, University of Oxford, Oxford OX2 6GG, United Kingdom}
\date{}  
\maketitle
\begin{abstract}

Traffic-induced failures, from packet loss in communication networks to congestion breakdown in transport systems, occur when flows progressively exhaust the edges they traverse. Path percolation models this process by removing edges along sampled origin--destination paths. Existing work assumes locally tree-like networks and deterministic shortest-path routing, leaving unclear how path degeneracy and routing stochasticity affect fragmentation in the clustered networks typical of real systems. We introduce a generalised path-percolation framework where paths are drawn from a temperature-controlled routing ensemble interpolating between geodesic and noisy transport. We argue based on box-covering renormalisation and our numerical experiments that, for any finite routing horizon $C$, the process coarse-grains to ordinary mean-field percolation. Routing details affect non-universal quantities, especially the percolation threshold $p_c$, through the entropy of the load distribution and the capacity of finite clusters to accommodate flow. Load entropy therefore acts as a robustness measure for networks under path-based failures.
When the routing horizon is tuned to the mean-field correlation length, $C=N^{1/3}$, within a source-uniform ensemble, the system enters a crossover regime with scaling exponents distinct from shortest-path percolation with infinite budget. In this regime, path elongation becomes decoupled in time from structural fragmentation: the characteristic path length reaches a growing maximum, associated with routing temperature, asymptotically ahead of the collapse of the giant component. These results clarify how microscopic routing organisation shapes macroscopic resilience, and identify path elongation as a measurable precursor of failure in communication and transport infrastructure. 
\end{abstract}

\section{Introduction}

Percolation is one of the paradigmatic models for studying phase transitions and critical phenomena \cite{stauffer_introduction_2018}. Originally proposed to study polymer gelation \cite{flory_constitution_1942}, percolation theory has since been successfully applied to many fields, including porous and composite materials \cite{morone_jamming_2019,bassett_extraction_2015,nabizadeh_network_2024}, social and economic systems \cite{cohen_breakdown_2001,motter_cascade-based_2002,cirigliano_neighbor-induced_2025,blagojevic_network_2024}, and epidemiology \cite{newman_spread_2002, moore_epidemics_2000,pastor-satorras_epidemic_2015}. Over the last two decades, standard bond and site percolation on uncorrelated, locally tree-like networks have been fully understood using message passing and generating functions \cite{newman_random_2001,newman_message_2023,cirigliano_scaling_2024}. However, many real-world processes are not governed by independent failures or node occupation. Specialised percolation models have therefore been proposed to capture the correlations observed in specific real-world systems and to assess their robustness \cite{artime_robustness_2024,li_percolation_2021,zhang_correlated_2019,makse_modeling_1998}.

A recent example is shortest-path percolation (SPP), in which edges are removed along shortest paths between randomly selected origin–destination pairs. This model captures resource exhaustion in demand-serving systems such as transportation and communication networks, and exhibits qualitatively different behaviour from classical percolation, including rapid dismantling of the giant component, homogenisation of the degree structure, and modified critical exponents. More broadly, the interplay between flow routing and network topology has been studied extensively across a wide range of systems, from communication networks and transportation infrastructure to biological and social networks \cite{meloni_traffic-driven_2009, arenas_communication_2001,lambiotte_networks_2019}. A central question is how the routing of flows — that is, the sampling of paths — interacts with the underlying topology, particularly under conditions of congestion, failures, and resource exhaustion \cite{guimera_optimal_2002,yan_efficient_2006, danila_optimal_2006}. SPP provides a minimal model for flow-induced network fragmentation, in which edges are removed indirectly through sequentially sampled paths, introducing a feedback between routing dynamics and network structure that is absent in ordinary bond percolation.

Existing studies of SPP have largely focused on locally tree-like networks, where the absence of clustering implies that shortest paths between node pairs are, with high probability, unique in the large-system limit. In contrast, real-world networks are typically highly clustered and contain multiple alternative routes between nodes. In such systems, routing is no longer uniquely determined, and the resulting path degeneracy can significantly alter macroscopic behaviour. This situation is reminiscent of Achlioptas processes, where small changes in the selection rule can qualitatively alter the nature of the transition due to the correlations and feedback introduced by the choice mechanism \cite{dsouza_explosive_2019}. While the impact of non-shortest-path routing has been explored in quantum communication \cite{hu_unveiling_2025} and traffic-driven spreading processes \cite{chen_traffic-driven_2022,yang_suppressing_2013}, it has not been systematically studied within a path-based percolation framework.

In this work, we address this question by introducing a model in which paths are sampled from a temperature-controlled ensemble that interpolates between geodesic and noisy routing. The ensemble is defined by two components: an origin–destination (OD) ensemble that determines how node pairs are selected, and a routing protocol whose temperature controls the degree of stochasticity. We show that for finite horizons, path-induced correlations are confined to a finite neighbourhood and the process renormalises to ordinary bond percolation. The routing temperature affects non-universal quantities, especially $p_c$ through the entropy of the edge-load distribution: concentrated load from low-temperature routing produces targeted, accelerated dismantling, whereas high-temperature routing homogenises edge removal and raises $p_c$. In the source-uniform ensemble, flow can additionally be absorbed by finite components, further shifting the transition relative to pair-uniform sampling. These results clarify the role of the path ensemble in shaping network robustness under flow-induced failures. More fundamentally, at the mean-field crossover horizon, we observe crossover exponent estimates, together with a growing pre-critical peak in the characteristic path length associated with the routing temperature. 

\section{Model}
Many real-world networks are not only small-world and scale-free but also highly clustered \cite{newman_networks_2018}. To understand the effect of path degeneracy, we use the Newman–Watts model to generate homogeneous and clustered percolation substrates \cite{newman_scaling_1999}. The Newman–Watts model is defined by three parameters $(N, K, \beta)$: the number of nodes $N$, the mean degree $K$ and the shortcut probability $\beta$. 
Each node is initially connected to its $K/2$ nearest neighbours on either side in a ring lattice; $\beta$ then determines the probability of adding a random long-range edge between each pair of nodes. This model produces networks with high clustering and short average path lengths while preserving a relatively homogeneous degree distribution. 

Our goal is to understand how the path ensemble, rather than only the underlying network topology, affects percolation. In ordinary bond percolation, edges are removed directly and independently. In path percolation, by contrast, edges are removed through sampled paths. The ensemble is determined by two components: an origin--destination (OD) ensemble and a routing protocol.

At step $t$, let $G_t=(V,E_t)$ be the remaining graph. An OD pair $(o,d)$ is first sampled from an ensemble
\begin{equation}
    Q_t(o,d),
\end{equation}
and then a path $\pi$ connecting $o$ and $d$ is sampled from
\begin{equation}
    R_t(\pi \mid o,d,G_t).
\end{equation}
The probability of sampling a path $\pi$ at step $t$ is therefore
\begin{equation}
    \mathbb{P}_t(\pi)
    =
    \sum_{o,d}
    Q_t(o,d)\,
    R_t(\pi\mid o,d,G_t).
    \label{eq:path_prob}
\end{equation}
The induced marginal probability that an edge $e$ is removed at step $t$ is
\begin{equation}
    \lambda_t(e)
    =
    \mathbb{P}_t(e\in \pi)
    =
    \sum_{o,d}
    Q_t(o,d)
    \sum_{\pi \ni e}
    R_t(\pi\mid o,d,G_t).
    \label{eq:edge_load}
\end{equation}
Path percolation thus defines a correlated edge-removal process. In general, $\lambda_t(e)$ is heterogeneous, reflecting the load carried by edge $e$ under the chosen OD and routing ensembles. Moreover, removals are correlated because all edges along the same sampled path are removed together. This distinguishes path percolation from bond percolation, where edge removals are independent and uniform.

\subsection{The global pair ensemble}
A natural reference ensemble is obtained by sampling origin--destination (OD) pairs uniformly from the set of admissible pairs within a finite horizon. We treat OD pairs as ordered pairs, so that \((o,d)\) and \((d,o)\) are distinct sampling events. At each removal event $t$, we define
\begin{equation}
A_t^{(C)} = \{\, (o,d) \in V_t \times V_t \;:\; d \in B_t^{(C)}(o),\; d \neq o \,\},
\end{equation}
where $V_t$ denotes the set of active nodes that have at least one neighbour, and
\begin{equation}
B_t^{(C)}(o) = \{\, d \in V_t : 1 \le d_{G_t}(o,d) \le C \,\},
\end{equation}
where $d_{G_t}(o,d)$ is the shortest-path distance in the current graph $G_t$. Hence, $B_t^{(C)}(o)$ is the set of nodes within C hops of $o$. The corresponding uniform OD ensemble is then given by
\begin{equation}
Q_t^{1}(o,d) =
\frac{1}{|A_t^{(C)}|}
\;\mathbf{1}\!\left\{(o,d)\in A_t^{(C)}\right\}, 
\label{eq:global_od}
\end{equation}
where $|A_t^{(C)}| = \sum_{o\in V_t} |B_t^{(C)}(o)|$.

\subsection{The local source ensemble} \label{sec:Source-uniform model}
In many physical, social, and technological systems, flows are not generated by uniformly selecting origin–destination pairs across networks. Instead, they are typically initiated from local sources and propagate through nearby connections. Examples include seed-borne diseases and information diffusion from users on social networks, where transmission originates from infected or active nodes and spreads through contact networks \cite{pastor-satorras_epidemic_2015,machado_effect_2022, makse_science_2024}. In such processes, interactions are activated locally, and the effective OD ensemble is naturally source-driven rather than globally uniform. Therefore, we consider the following source-uniform ensemble. 

At each step $t$, we choose an origin node $o$ uniformly from $V_t$. The destination $d$ is then sampled uniformly from its C-hop neighbourhood obtained by the Breadth-First Search (BFS). The corresponding OD ensemble is
\begin{equation}
    Q_t^{2}(o,d)
    =
    \frac{1}{|V_t|}
    \frac{\mathbf{1}\{d\in B_t^{(C)}(o)\}}
         {|B_t^{(C)}(o)|}.
    \label{eq:local_od}
\end{equation}

This defines a source-uniform local ensemble: each origin is sampled with equal probability, and the destination is chosen uniformly among nodes reachable from that origin within C hops. If instead origins are sampled with weights proportional to the size of their BFS balls, $|B_t^{(C)}(o)|$, the resulting sampling becomes equivalent to the pair-uniform ensemble. 

\begin{equation}
    Q_t^{1}(o,d)
    = \frac{|B_t^{(C)}(o)|}{\sum_{o\in V_t} |B_t^{(C)}(o)|}\frac{\mathbf{1}\{d\in B_t^{(C)}(o)\}}
         {|B_t^{(C)}(o)|}
\end{equation}

This distinction is particularly important in heterogeneous networks, where high-degree nodes generate large BFS balls and are therefore overrepresented under pair-uniform sampling. As a result, pair-uniform ensembles tend to emphasise hub-dominated traffic patterns, contributing to the homogenisation effects observed in~\cite{kim_shortest-path_2026}.

\subsection{The horizon C}
In many physical, social, and technological systems, such as information spreading and transportation, interactions are not global: signals, packets, or agents typically propagate within a limited range, and the probability of interaction decreases with distance \cite{lopez_limited_2007,li_percolation_2011}. The parameter C controls the reachability of origin–destination (OD) pairs and determines the effective ensemble of OD pairs. When C is small, the process is local and strongly constrained by network geometry. As C increases, the accessible region grows. When C equals the network diameter, the source-uniform ensemble is identical to the pair-uniform ensemble within a fixed connected component. We choose C to be at least comparable to the characteristic distance of the original network, so that the explored subgraph reflects its large-scale structure. In homogeneous small-world networks, this corresponds to the average distance between ends of the shortcuts $1/(\beta K)$, ensuring that the local neighbourhood already exhibits small-world properties \cite{newman_scaling_1999}. 

The horizon $C$ is a natural measure of correlation length in networks. For fixed $C$, each removal event only couples edges inside a finite BFS neighbourhood. Such a neighbourhood can be covered by a box of size
$2C+1$ in the burning algorithm \cite{song_self-similarity_2005,song_how_2007} developed for network renormalisation. At scales larger than this box size, the path-induced correlations are expected to become irrelevant, and the process only perturbs the mean-field fixed point. The situation changes when $C$ increases with $N$ as reported in recent works \cite{kim_shortest-path_2024,kim_shortest-path_2026,meng_path_2025}. In this regime, path-induced correlations are no longer finite-range perturbations,
and a crossover away from the finite-$C$ scaling behaviour can occur. We use this distinction between fixed and growing horizons to separate local path-removal effects from genuinely critical-scale path correlations.

\subsection{Temperature-controlled routing protocol}

Given an OD pair $(o,d)$, we sample a path using a temperature-controlled preferential walk. The walk is performed from $d$ back to $o$ on the current graph $G_t$, using the BFS distances from the origin $o$ as an energy landscape. If the walker is currently at node $c$ and $u$ is a neighbour of $c$ along an unremoved edge, define
\begin{equation}
    \Delta l = d_o(c)-d_o(u),
\end{equation}
where $d_o(u)$ denotes the shortest-path distance from $o$ to $u$ in $G_t$. We restrict the moves to $\Delta l \in \{0,1\}$, so that the walk is strictly non-ascending in the distance landscape. The transition weight is
\begin{equation}
    w(c\to u)
    =
    \exp\!\left(\frac{\Delta l}{T}\right),
    \label{eq:path_weight}
\end{equation}
where $T$ is the routing temperature: $T\to 0$ recovers geodesic routing along BFS shortest paths, while $T\to\infty$ samples uniformly among non-ascending moves. The next node is drawn with probability proportional to the transition weight, subject to the constraint that no node is visited twice (the walk is self-avoiding). Note that the walk cannot be trapped: every node at distance $k>0$ from $o$ has at least one neighbour at distance
$k-1$, which cannot have been visited previously since all visited nodes lie at distance $\geq k$. A descending move is therefore always available, and the walk terminates at $o$ in at most finitely many steps.

In the low-temperature limit, moves that decrease the distance to the origin are strongly
favoured, and the sampled path is close to a geodesic. In the high-temperature limit,
horizontal moves and detours become more likely, so the path ensemble includes a broader
set of non-shortest paths. The parameter $T$ therefore interpolates between a
shortest-path ensemble and a noisy routing ensemble. This construction does not
enumerate all non-shortest paths, which remains computationally difficult. Instead, it assigns
a dynamical probability measure to a subset of self-avoiding paths, in the same spirit as other path-sampling studies based on random walk \cite{francoisse_bag--paths_2017,newman_measure_2005}. Equivalently, the walker moves
on a discrete energy landscape whose basin is located at the origin node $o$.

After a path is sampled, all edges on the path are removed from the graph. Repeating
this procedure generates an ordered edge-removal sequence. The percolation observables are
then computed by reversing this sequence and using a Newman--Ziff-type reconstruction procedure. The implementation details are covered in the Supplementary Materials.

\begin{figure}[!t]
    \centering
    \centering
    \includegraphics[width=\linewidth]{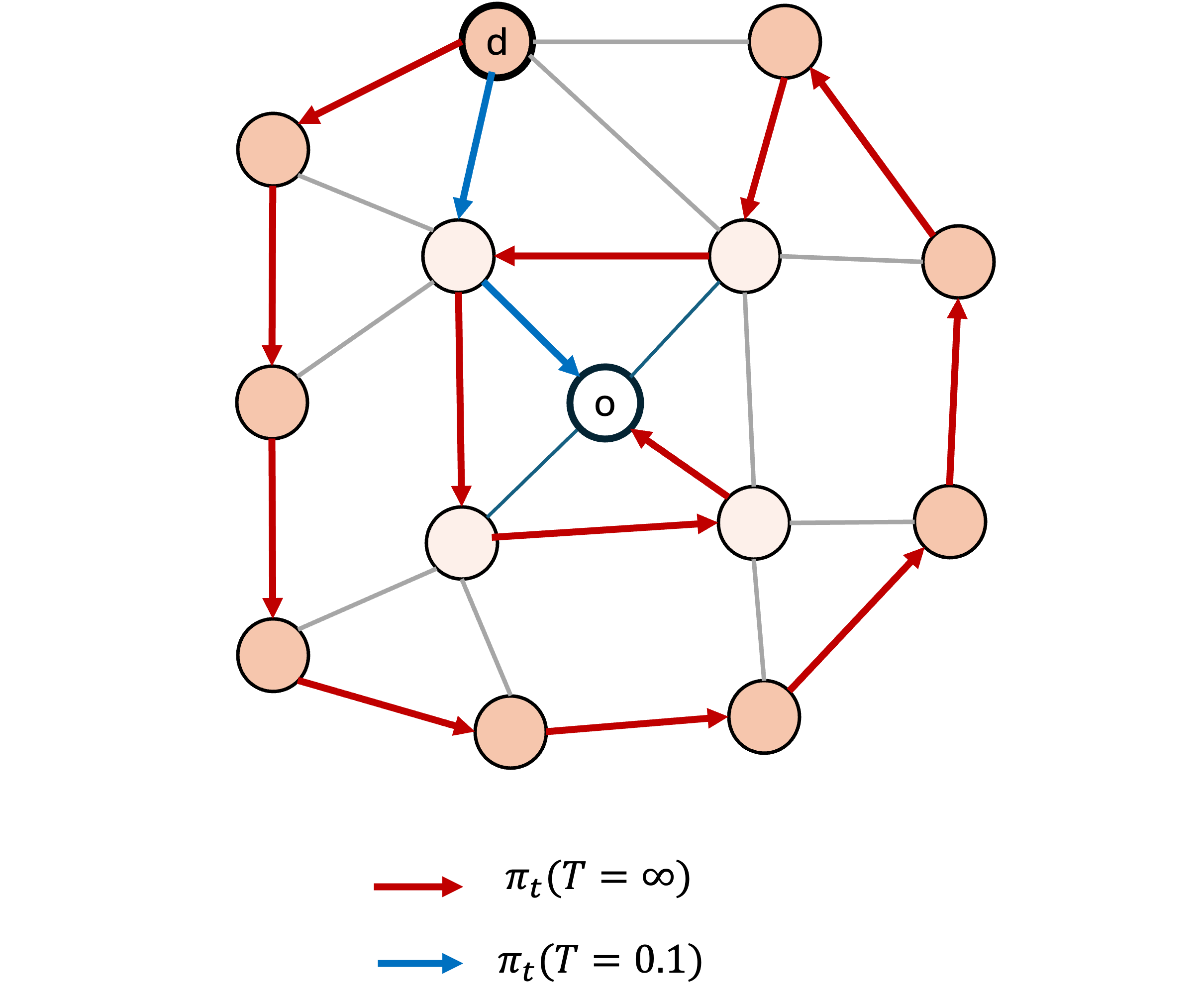}
    \caption{A typical subnetwork (BFS ball) returned from the breadth-first search with two hops ($C = 2$). The distance between nodes and the origin is transparency-coded. The sampled shortest and noisy path are denoted by blue and red edges, respectively.}
    \label{fig:Schematic1}
\end{figure}

\clearpage
\newpage
\section{Results}
\subsection{Path ensemble with finite horizon}
\begin{figure}[!ht]
    \centering
    \begin{subfigure}{0.9\textwidth}
        \includegraphics[width=\linewidth]{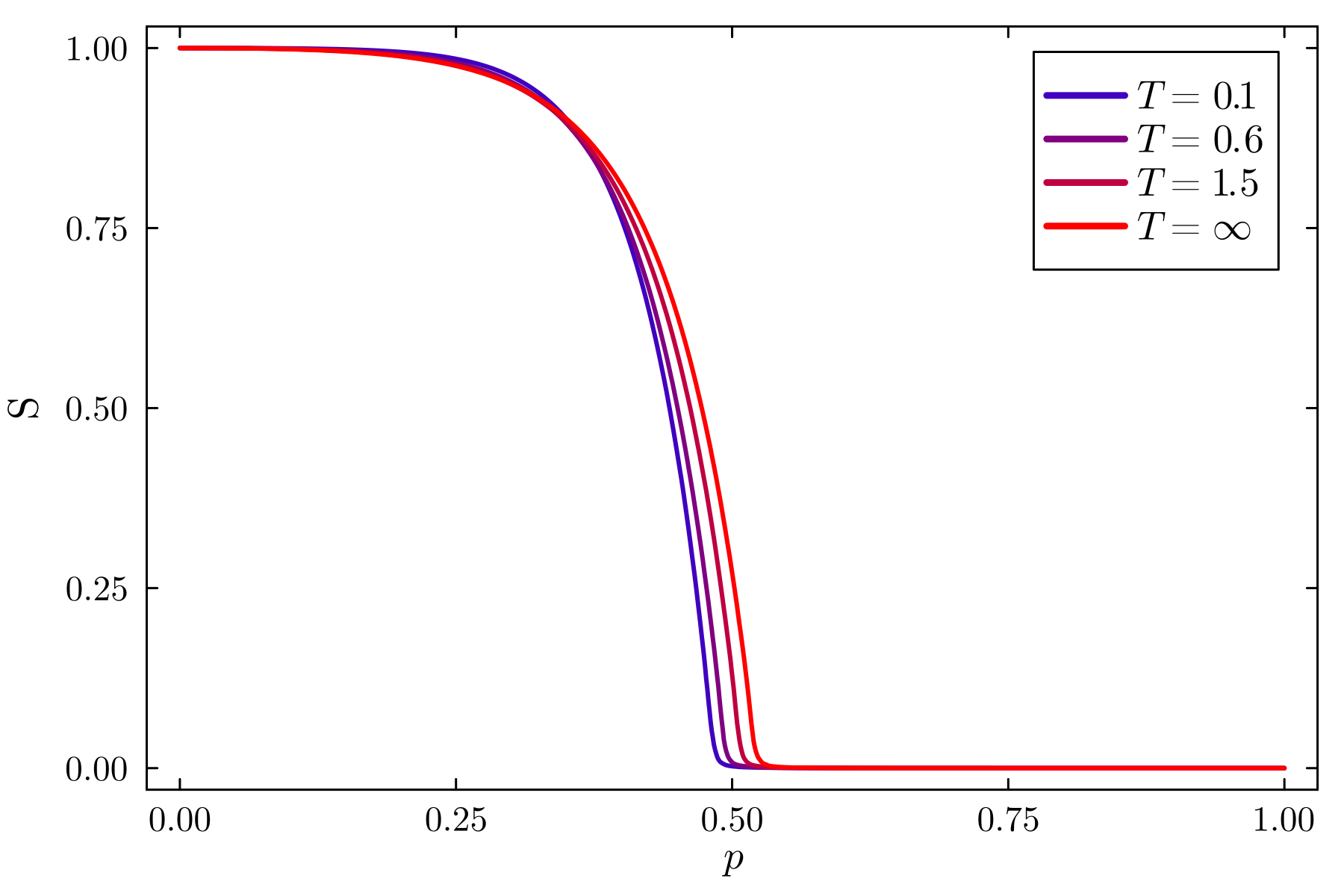}
        \caption{Path percolation in pair-uniform ensemble with finite horizon in the Newman-Watts model with $2^{21}$ nodes, $K=4$, $\beta=0.1$. We set $C = 3$ across different network sizes, which is greater than the typical distance of shortcuts $\frac{1}{\beta K} = 2.5$.}
        \label{fig:pair-uniform_phase_diagram}
    \end{subfigure}
    \hfill
    \begin{subfigure}{0.9\textwidth}
        \includegraphics[width=\linewidth]{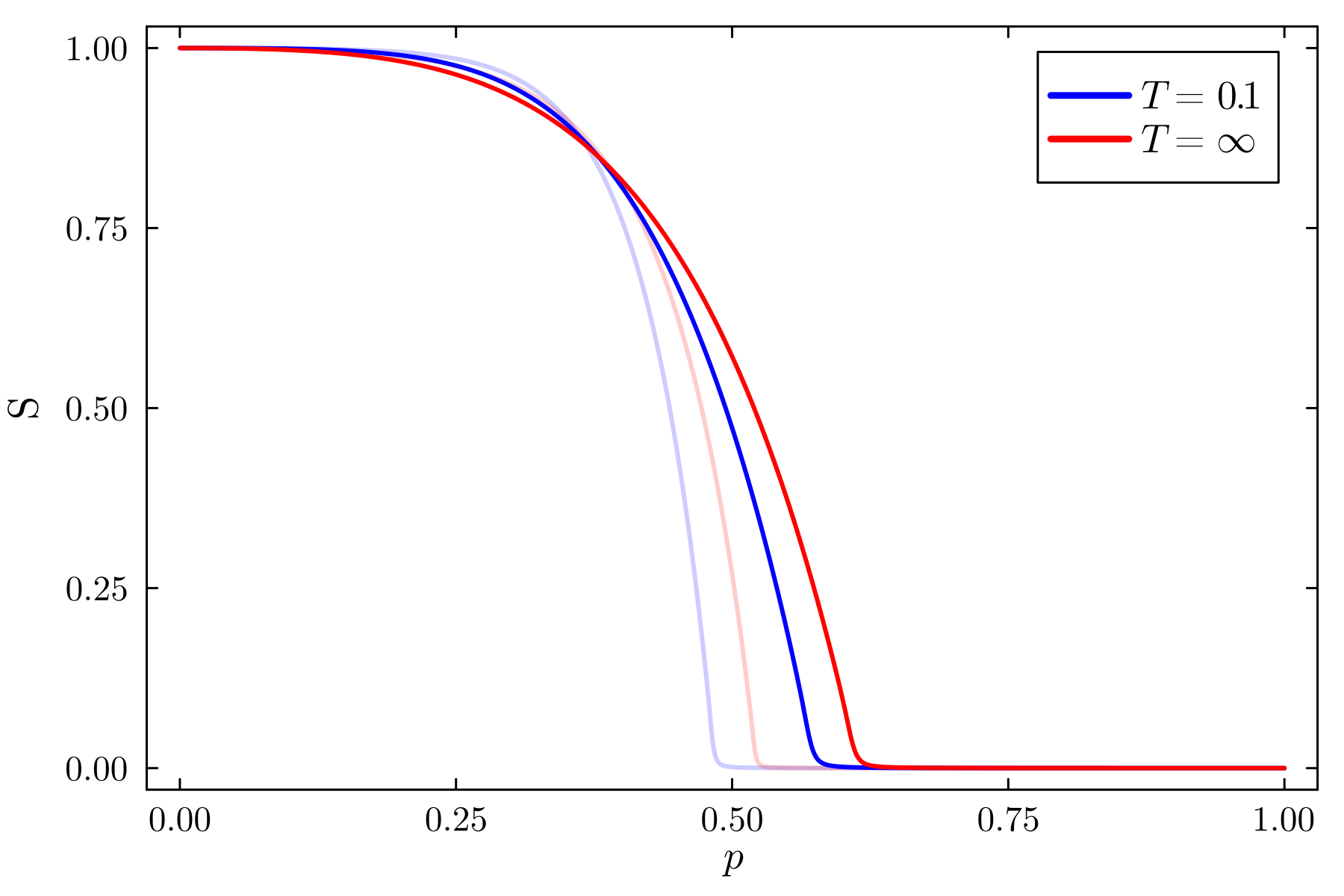}
        \caption{Path percolation in the source-uniform ensemble with finite horizon on the same network substrate. The transparent lines denote the results from the path percolation in the pair-uniform ensemble.}
        \label{fig:source-uniform_phase_diagram}
    \end{subfigure}
    \caption{Finite-horizon path percolation on Newman--Watts networks. Higher routing temperature shifts the transition to larger \(p\), indicating that noisy routing delocalises path-based damage. In the source-uniform ensemble, finite clusters can also accommodate local demand, further increasing robustness.}
    \label{fig:PhaseDiagram}
\end{figure}

We first investigate the impact of the path ensemble with a finite horizon. In \Cref{fig:PhaseDiagram}, we plot the relative size of the giant component against the fraction of the edges removed with different routing temperatures in pair-uniform and source-uniform ensembles. For the pair-uniform ensemble at the low-temperature region ($T \le 0.2$, shown in the Supplementary Materials), the process reduces to the shortest-path percolation (SPP), with C playing the role of the budget \cite{kim_shortest-path_2024}. As the temperature increases, path sampling becomes progressively less constrained, allowing for the exploration of suboptimal routes. A clear trend emerges: increasing the routing temperature systematically shifts the critical point $p_c$ to larger values, indicating that the network becomes more robust against path-based removal. This effect is observed in heterogeneous substrates (See Supplementary Materials), suggesting that it is not tied to a specific degree distribution but rather to the routing protocol itself. 

Source-uniform ensembles model transmission initiated from local seeds: each origin is sampled uniformly, while destinations are restricted to the $C$-hop neighbourhood of the source. Since this sampling does not explicitly favour nodes with large BFS balls---which in heterogeneous networks are typically associated with hubs---one might expect the source-uniform and pair-uniform ensembles to exhibit similar transition behaviour in the Newman--Watts model, whose degree distribution is homogeneous. However, surprisingly, the percolation threshold in the source-uniform ensemble is systematically shifted to larger $p$ relative to its pair-uniform counterpart across the temperature spectrum (\Cref{fig:source-uniform_phase_diagram}). Flows initiated from local seeds are thus intrinsically less disruptive than globally sampled pair interactions, even in the absence of strong degree heterogeneity. The origin of this effect lies in finite connected clusters. In the pair-uniform ensemble, the probability of selecting two nodes in the same finite cluster vanishes in the thermodynamic limit whenever a giant component exists, so asymptotically every sampled path consumes giant-component edges. In the source-uniform ensemble, by contrast, a finite fraction $1-S$ of flows originates in finite clusters; the associated removals are confined there, sparing the giant component and delaying its collapse. The difference reflects that connectivity alone does not fully characterise network functionality: communication or transport may remain feasible within locally connected regions without participation in the giant component~\cite{hamedmoghadam_percolation_2021}. 

To understand the temperature-dependent shift of the percolation point, we consider path percolation as a flow-induced dismantling process where each removal corresponds to choosing a packet and destroying the path it traverses. We measured the edge-load distribution generated by the path ensemble. At fixed removed-edge fractions, we sampled paths at different routing temperatures before the next damage and recorded the number of visits to each edge $(i,j)$, $\sigma_{ij}$. We define the normalised load distribution $p_{ij} = \sigma_{ij} / \sum_{i<j} \sigma_{ij}$. This quantity provides a Monte Carlo estimate of the edge load, which is proportional to the marginal edge usage probability $\lambda_t(e)$ in \Cref{eq:edge_load}, up to a weighting of path length. 

\begin{figure}[!t]
    \centering
    \begin{subfigure}{\textwidth}
        \centering
        \includegraphics[width=\linewidth]{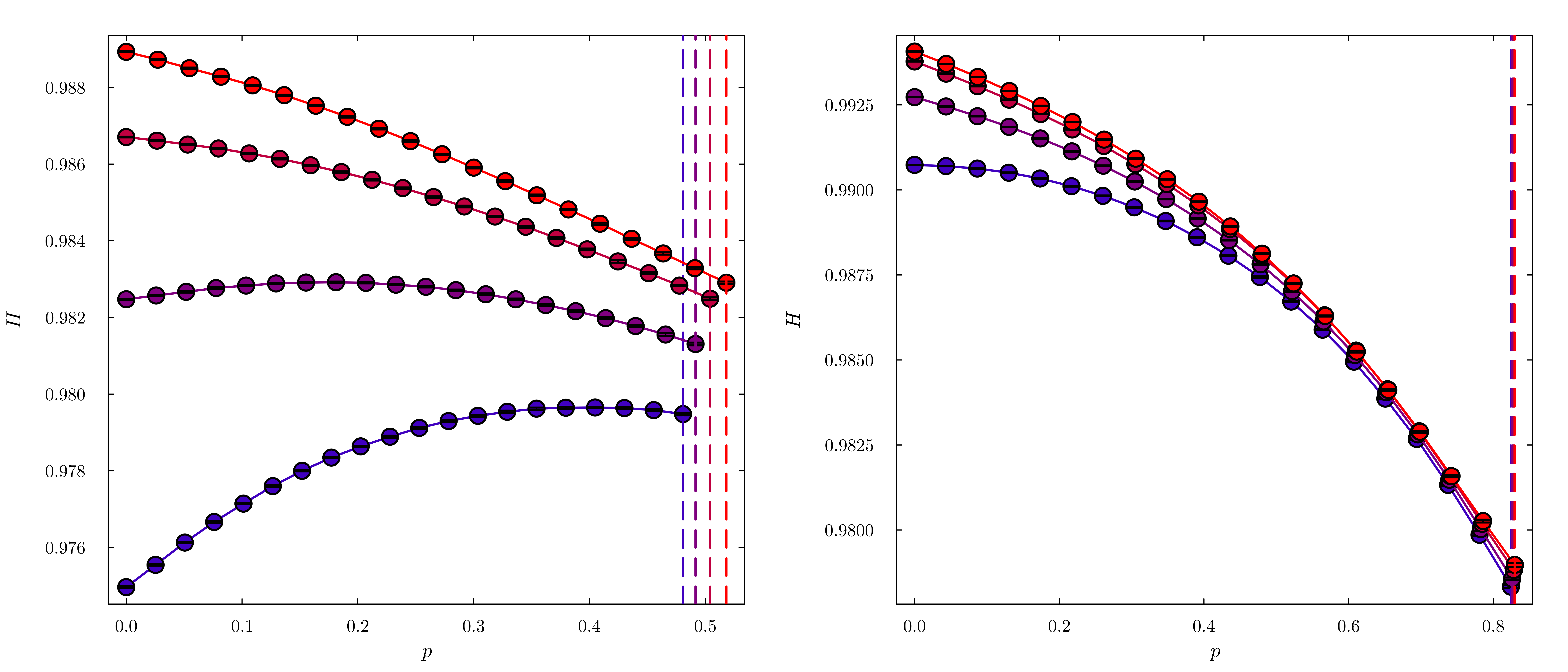}
        \caption{The evolution of entropy as a function of $p$ for Newman-Watts models with $\beta = 0.1$ (left) and $1$ (right). Different curves denote routing temperatures used in the phase diagram (\Cref{fig:PhaseDiagram}). The vertical lines denote the estimated percolation point.}
        \label{fig:snapshot_entropy_comparsion}
    \end{subfigure}
    
    \begin{subfigure}{\textwidth}
        \centering
        \includegraphics[width=\linewidth]{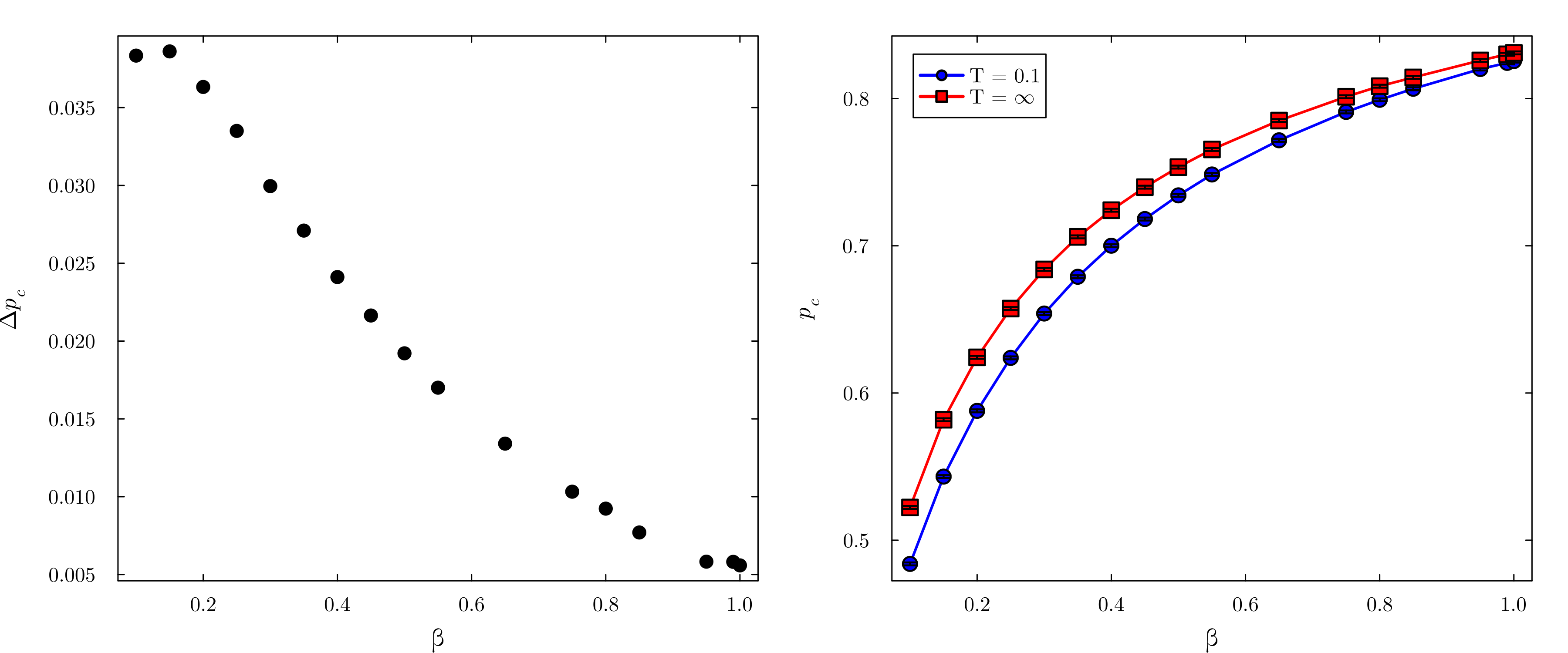}
        \caption{The difference in the critical point estimated by the finite-size scaling analysis plotted as a function of $\beta$.}
        \label{fig:pc_beta}
    \end{subfigure}
   \caption{Load delocalisation explains the temperature-dependent shift of the percolation threshold. Panel (a) shows the normalised load entropy \(H\) as a function of the removed-edge fraction \(p\). Panel (b) shows that the separation between low- and high-temperature critical points decreases as the shortcut probability \(\beta\) increases.}
    \label{fig:entropy_effect}
\end{figure}

We quantify the localisation of this load distribution using the normalised entropy
\begin{equation}
H = -\left(\frac{\sum_{i<j} p_{ij} \log p_{ij}}{\log(M)}\right),
\end{equation}
where \(M\) is the total number of edges. 
The entropy increases with \(T\), as shown in \Cref{fig:snapshot_entropy_comparsion}, demonstrating a difference in traffic delocalisation. At low temperatures, shortest-path routing concentrates flow on a small subset of high-betweenness edges, which are strongly correlated with hubs in heterogeneous networks \cite{goh_load_2005}. Removing such paths fragments the network rapidly. At high temperatures, by contrast, the load is spread
over a larger fraction of the edge set, making the damage process less targeted and shifting \(p_c\) upward.

With this intuition, we further vary the network topology by changing the shortcut probability $\beta$ in the Newman--Watts model. As shown in \Cref{fig:pc_beta}, the difference between the critical points at low and high temperature decreases as $\beta$ increases. For small $\beta$, the number of alternative routes is limited, so routing produces localised loads, and the entropy difference between low- and high-temperature routing is large. For $\beta=1$, shortest paths are highly degenerate and distributed across the network; consequently, the entropy difference between routing protocols becomes small, and the corresponding shift in $p_c$ nearly vanishes (\Cref{fig:snapshot_entropy_comparsion}).

This observation suggests that increasing the availability of alternative routes reduces the sensitivity of the percolation threshold to the routing noise. By the same reasoning, increasing the routing horizon $C$ may also diminish the impact of routing noise by enlarging the accessible path ensemble. Therefore, topology and routing influence network robustness through a common mechanism: traffic delocalisation. We therefore identify the entropy of the load distribution as a natural measure for robustness under path-based failures. We have also verified in the Supplementary Materials that the participation ratio provides a consistent measure of traffic localisation and leads to the same qualitative conclusions.

For the critical exponents, we focus on the scaling of the order parameter $S_c(N) $, the susceptibility $\chi_c(N)$, and the finite-size shift of the critical point $p_c(N)$, using the standard finite-size scaling ansatz and the event-based ensemble \cite{li_explosive_2023, fan_universal_2020}. The results are summarised in \Cref{tab:exponents}. At both low and high temperatures, the estimated critical exponents are consistent with the mean-field universality class, which is expected for the Newman-Watts model, where classical bond percolation exhibits mean-field critical behaviour \cite{newman_scaling_1999}. This observation can be understood phenomenologically from a network renormalisation perspective. For a finite horizon $C$, the path sampling introduces correlations in edge removal that are restricted to a local neighbourhood. Specifically, each breadth-first search region explored during path construction can be coarse-grained into a supernode of box size $l_b = 2C+1$.  Correlations from the path ensemble are irrelevant at scales larger than $l_b$, and the system converges to the mean-field limit. 

\clearpage
\newpage
\subsection{Crossover scaling and the decoupling of path elongation from
structural fragmentation}
\label{sec:crossover}

A key question is whether the critical scaling of the source-uniform ensemble crosses over to the one observed in the shortest path percolation with infinite budget when the routing horizon is tuned to the intrinsic critical scale of the network. In mean-field percolation, critical clusters have mass $s_c\sim N^{2/3}$, and their chemical distance scales as $l_\xi\sim s_c^{1/2}$. Hence $l_\xi\sim N^{1/3}$. Setting $C = A\,N^{1/3}$ probes the crossover regime in which path-induced correlations are no longer finite-range perturbations.

\begin{figure}[!ht]
    \centering

    \begin{subfigure}{\textwidth}
        \centering
        \includegraphics[width=\linewidth]{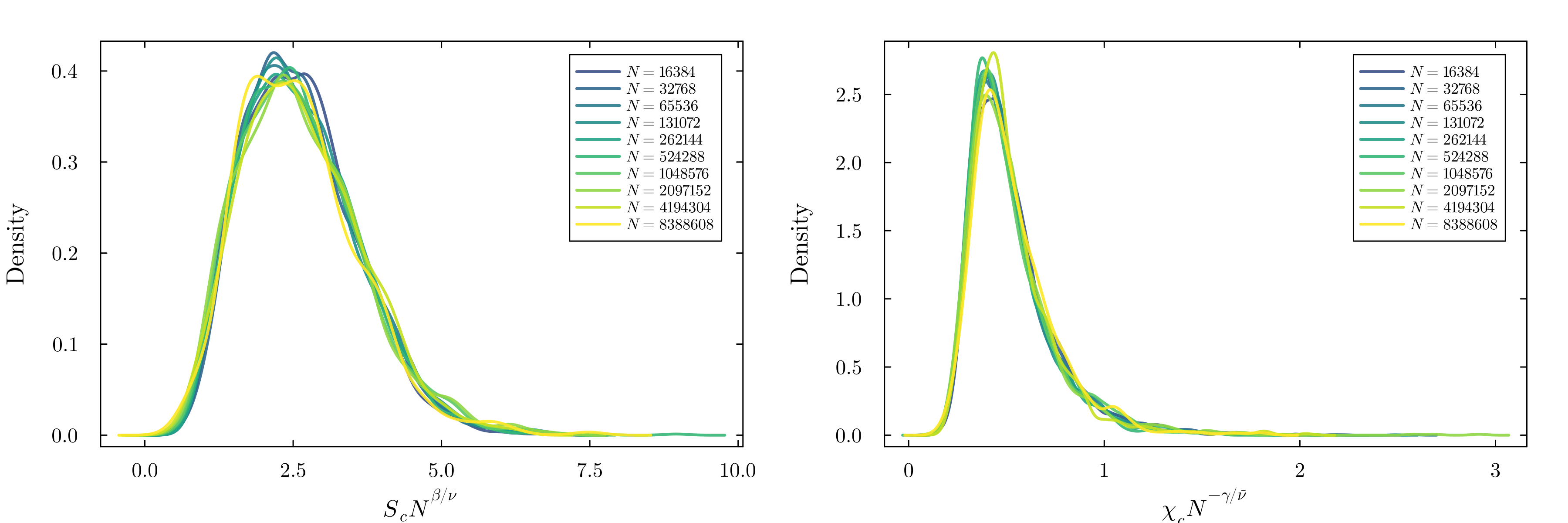}
        \caption{$T=\infty$}
        \label{fig:FSS_collapse_highT}
    \end{subfigure}

    \vspace{0.5em}

    \begin{subfigure}{\textwidth}
        \centering
        \includegraphics[width=\linewidth]{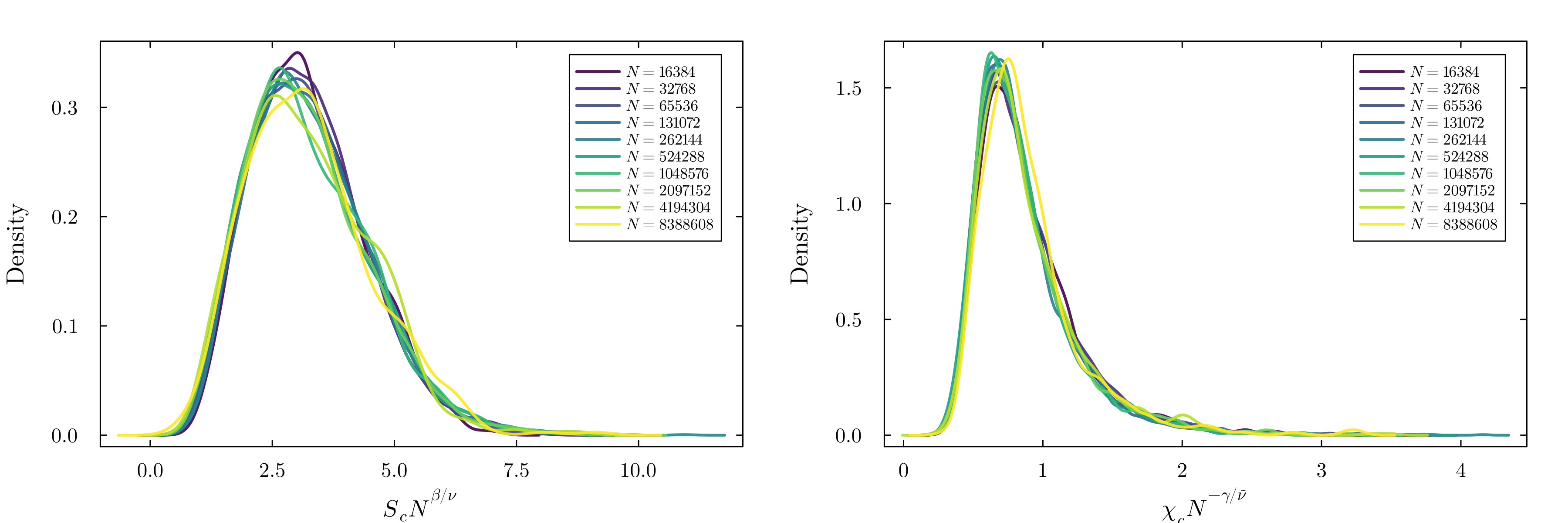}
        \caption{$T=0.1$}
        \label{fig:FSS_collapse_lowT}
    \end{subfigure}

\caption{Collapse of the probability distributions of the rescaled order parameter \(S_c\) and susceptibility \(\chi_c\) for high- and low-temperature routing in the source-uniform ensemble with horizon \(C=N^{1/3}\).}
    \label{fig:FSS_collapse}
\end{figure}

The resulting exponent estimates differ from both the mean-field values and those reported for shortest-path percolation with an infinite budget, while the data collapse remains robust across both temperature regimes (\Cref{fig:FSS_collapse}). Each ingredient of the path ensemble leaves its own imprint on the estimates. Increasing the prefactor $A$ in $C = A\, N^{1/3}$ shifts the exponents monotonically towards those of the saturated horizon $C=N$, so the routing horizon controls the overall drift away from mean-field behaviour. At fixed horizon, the routing temperature modifies the susceptibility exponent, whereas the giant-component exponent is comparatively insensitive. The demand dispatch matters as well: at $C=N$ we recover $\beta/\bar{\nu} \approx 0.21$, but $\gamma/\bar{\nu}$ remains smaller than the value reported for pair-uniform shortest-path percolation \cite{kim_shortest-path_2024}, reflecting that finite clusters in the source-uniform ensemble absorb local demand and thereby reduce the divergence of the cluster-size fluctuations. The crossover estimates also deviate systematically from the hyperscaling relation $2\beta/\bar{\nu}+\gamma/\bar{\nu}=1$, with the sum decreasing as the horizon increases. Taken together, these observations indicate that the critical exponents depend primarily on the horizon, i.e., the range of path-induced correlations, but are also sensitive to the demand dispatch and the routing temperature. All results are summarised in \Cref{tab:exponents}.

\begin{figure}[!ht]
    \centering

    \begin{subfigure}{\textwidth}
        \centering
        \includegraphics[width=\linewidth]{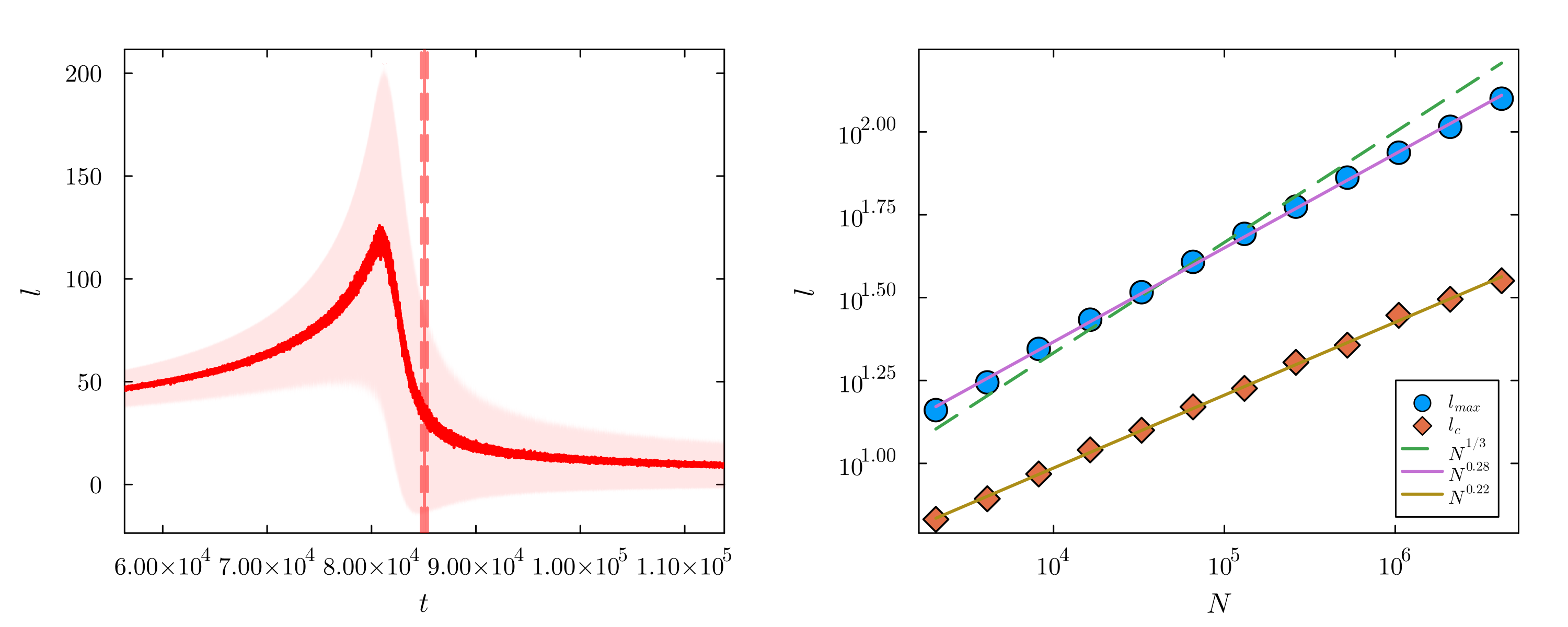}
        \caption{$T = \infty$}
        \label{fig:path_highT}
    \end{subfigure}

    \vspace{0.5em}

    \begin{subfigure}{\textwidth}
        \centering
        \includegraphics[width=\linewidth]{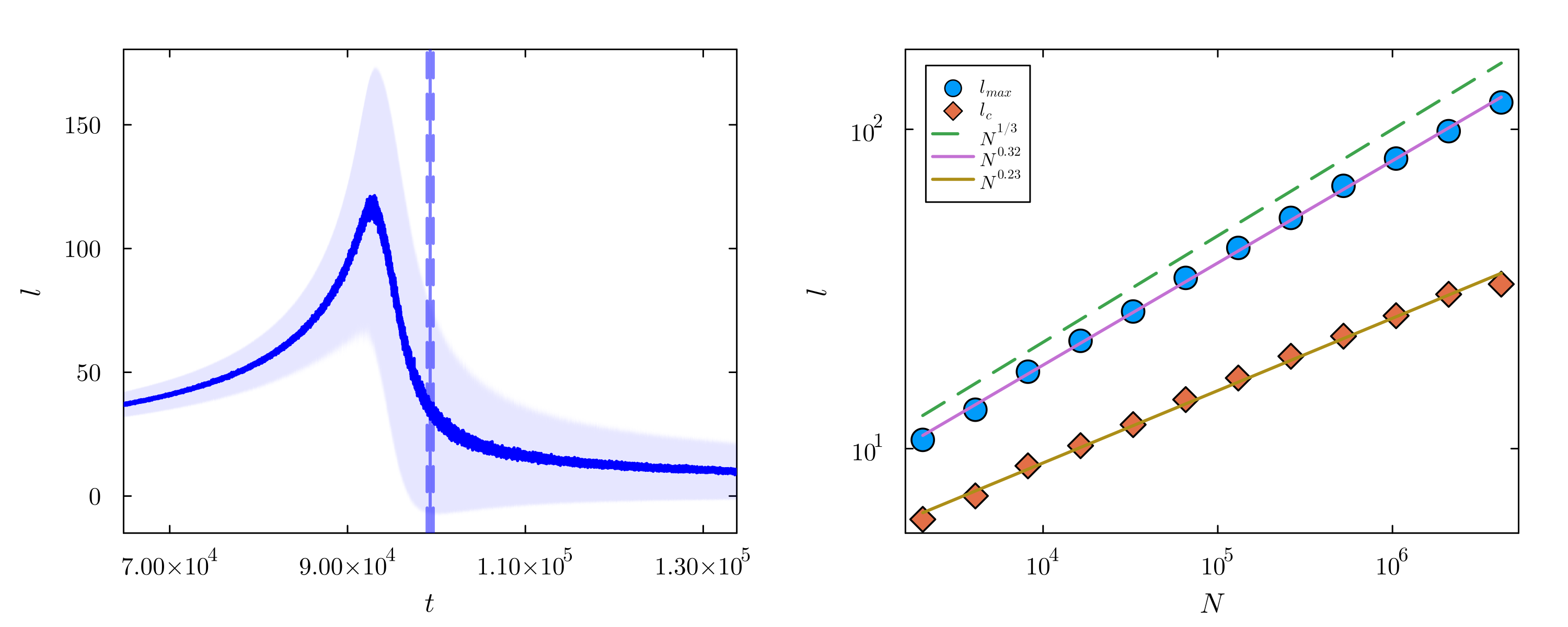}
        \caption{$T = 0.1$}
        \label{fig:path_lowT}
    \end{subfigure}

    \caption{
    Left: fragmentation dynamics on the Newman-Watts network with $2^{22}$ nodes. The vertical lines denote the trial that triggers the collapse of the giant component. The ribbon encloses the region within one standard deviation. Right: the scaling of $l_{max}$ and $l_c$ with N in the source-uniform ensemble with a growing routing horizon $C=N^{1/3}$.}
    \label{fig:path_scaling_crossover}
\end{figure}

The most striking consequence of this crossover ($C = N^{1/3}$) is a dynamical separation between two processes that coincide in ordinary percolation: \emph{path elongation} and \emph{structural fragmentation}. As shown in \Cref{fig:path_scaling_crossover}, the average sampled path length $l(t)$ grows and reaches a maximum $l_{\max}$ strictly before the collapse of the giant component at $t_c$. This pre-critical peak is not merely a finite-size effect; the data show that it escapes the critical time window entirely in the large-$N$ limit.

\begin{figure}[!ht]
    \centering

    \begin{subfigure}[b]{0.48\textwidth}
        \centering
        \includegraphics[width=\linewidth]{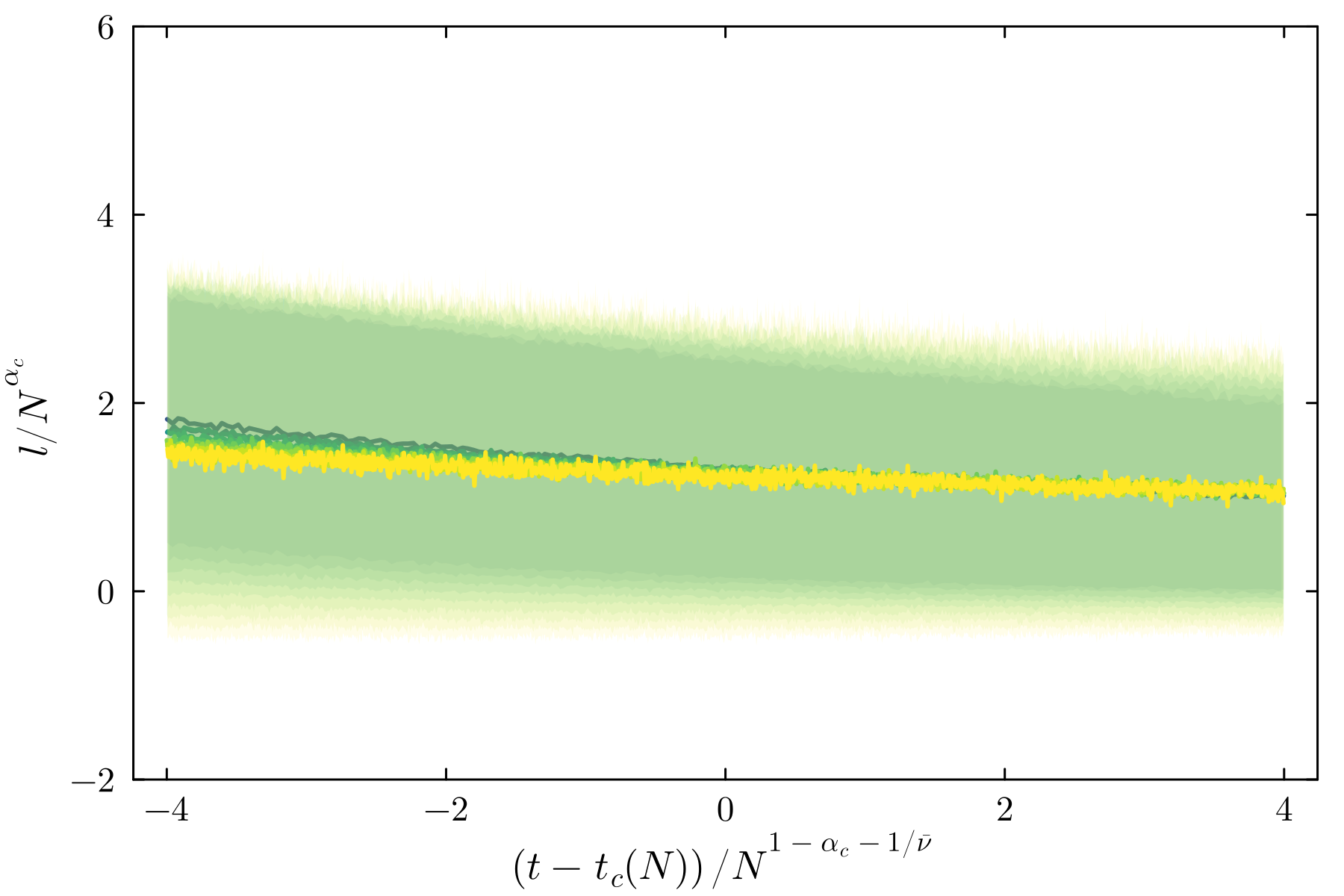}
        \caption{$T = \infty$}
        \label{fig:path_collapse_highT}
    \end{subfigure}
    \hfill
    \begin{subfigure}[b]{0.48\textwidth}
        \centering
        \includegraphics[width=\linewidth]{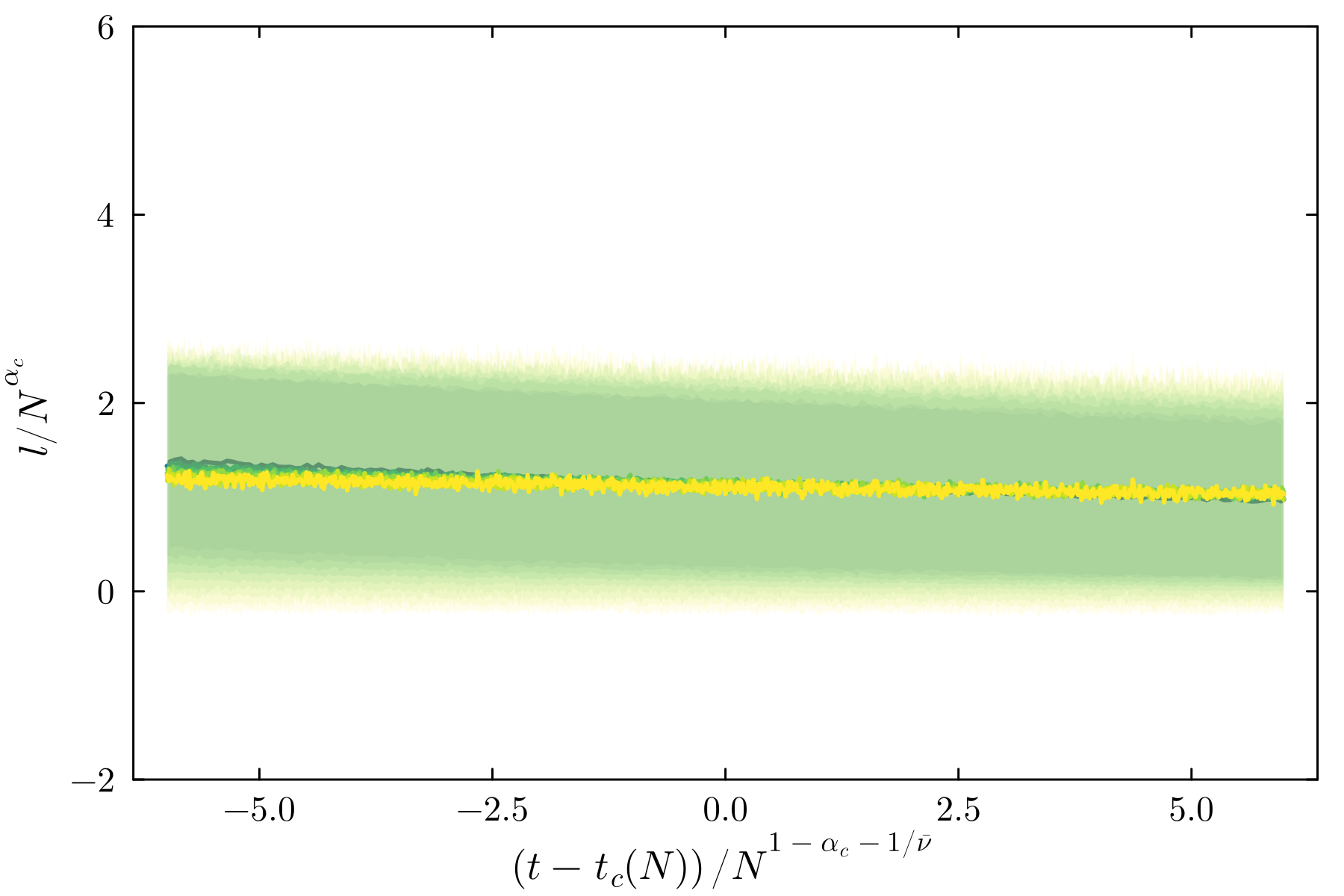}
        \caption{$T = 0.1$}
        \label{fig:path_collapse_lowT}
    \end{subfigure}

    \caption{Data collapse of the sampled path length and its standard deviation around the critical time $t_c(N)$ for high- and low-temperature routing in the source-uniform ensemble with horizon $C = N^{1/3}$. The near-constancy of the scaling function $G$ shows that $l$ varies only weakly within the critical window of width $N^{1-a_c-1/\bar\nu}$, which underpins the conclusion that the pre-critical peak in $l$ lies outside this window.}
    \label{fig:path_collapse_crossover}
\end{figure}

Each removal event deletes the edges of one sampled path, so the removed-edge fraction follows the dynamics,
\begin{equation}
    \frac{dp}{dt} = \frac{l(t,N)}{M},
    \label{eq:dpdt}
\end{equation}
where $M \propto N$ is the total number of edges. Unlike ordinary percolation, where $p$ and $t$ are proportional, the conversion factor here is the instantaneous path length itself. Near the critical point, the path length scales as $l_c \sim N^{a_c}$, so a deviation in the control parameter maps to a deviation in event time as
\begin{equation}
    |t - t_c(N)| \sim \frac{M}{l_c}\,|p - p_c(N)| \sim N^{1-a_c}\,|p - p_c(N)|.
\end{equation}
The structural critical window has width $|p - p_c(N)| \sim N^{-1/\bar\nu}$, which translates into a time window of width $N^{1-a_c-1/\bar\nu}$. The finite-size scaling of the path length is therefore naturally formulated in the shifted, rescaled time variable
\begin{equation}
    l(t,N) = N^{a_c}\, G\!\left(\frac{t - t_c(N)}{N^{1-a_c-1/\bar\nu}}\right).
\end{equation}\label{eq:fss_ell}
As shown in \Cref{fig:path_collapse_crossover}, the scaling function $G$ is nearly constant, indicating that $l$ varies only weakly within the critical window of width $N^{1-a_c-1/\bar\nu}$. The value of $1-a_c-1/\bar\nu$ is close to 0.33 in both temperature regimes, consistent with the imposed crossover scale $C = N^{1/3}.$

\begin{figure}[htbp]
    \centering
    
    \begin{subfigure}[b]{0.48\textwidth}
        \centering
        \includegraphics[width=\linewidth]{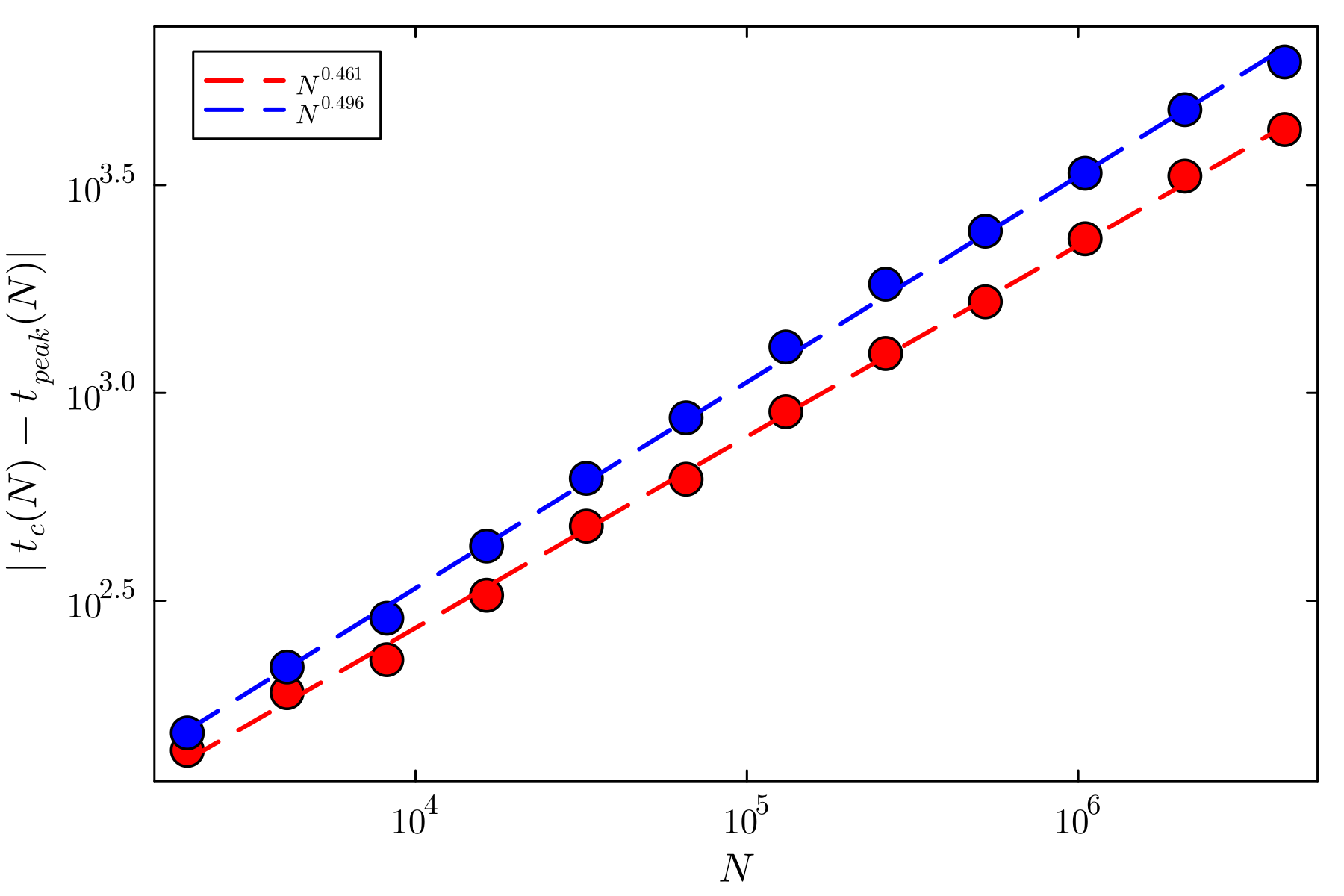}
        \caption{}
        \label{fig:tctpeak}
    \end{subfigure}
    \hfill
    \begin{subfigure}[b]{0.48\textwidth}
        \centering
        \includegraphics[width=\linewidth]{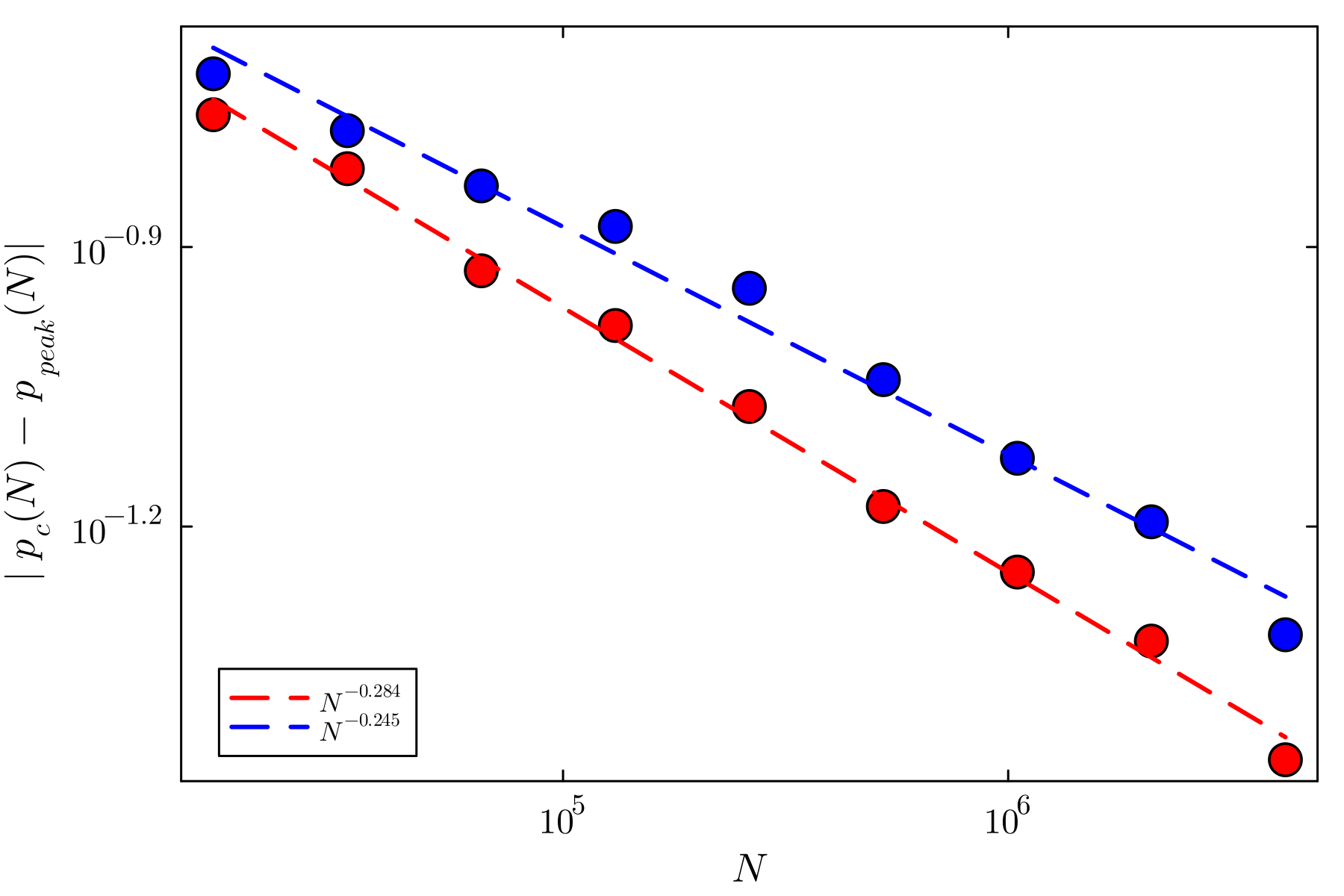}
        \caption{}
        \label{fig:pcvsppeak}
    \end{subfigure}
    
   \caption{Finite-size scaling of the separation between the structural collapse point and the peak in path length. Panel (a) shows \(|t_c-t_{peak}|\); panel (b) shows \(|p_c-p_{peak}|\). Results are shown for low- and high-temperature routing.}
    \label{fig:criticalwindow}
\end{figure}

The separation between the path-length peak and the structural critical time grows as
\begin{equation}
    |t_c(N) - t_{\mathrm{peak}}(N)| \;\sim\;
    \begin{cases}
        N^{0.50} & T = \infty, \\[4pt]
        N^{0.46} & T = 0.1,
    \end{cases}
    \label{eq:peak_separation}
\end{equation}
(\Cref{fig:tctpeak}). Since both exponents exceed $1 - a_c - 1/\bar\nu \approx 0.33$, the peak lies strictly outside the critical window for large $N$: path elongation and giant-component collapse become asymptotically distinct events on the time scale.

\begin{figure}
    \centering
    \includegraphics[width=1\linewidth]{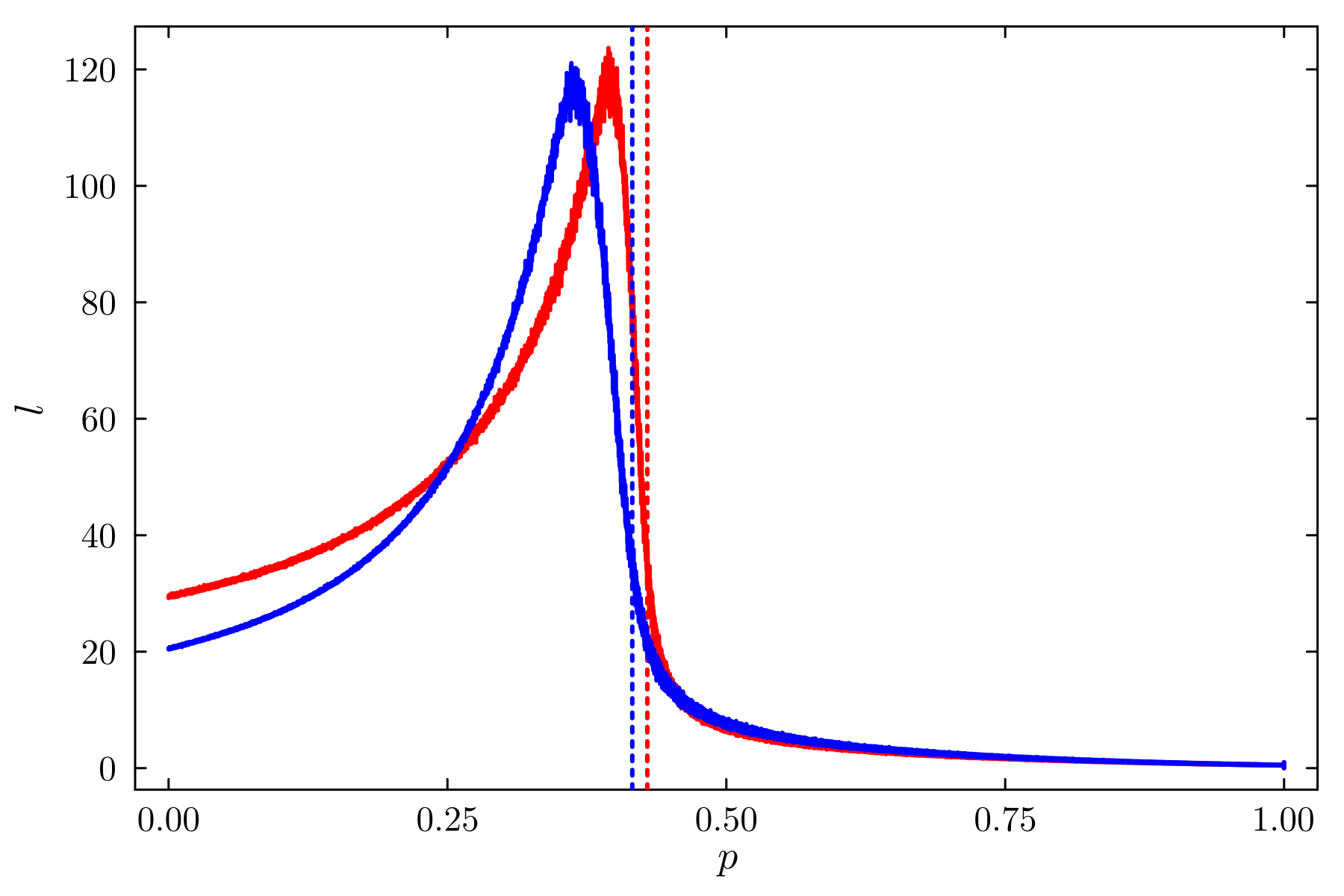}
    \caption{The sampled average path length against the fraction of removed edges, using \Cref{eq:dpdt} with $N = 2^{22}$ nodes. The vertical lines denote the critical point.}
    \label{fig:lvsp}
\end{figure}

Translating to the control parameter $p$ via \Cref{eq:dpdt}, we calculate $l(p)$ in \Cref{fig:lvsp}. The peak location $p_{peak}(N)$ converges to $p_c(N)$ with the following scaling, 
\begin{equation}
    |p_c(N) - p_{peak}(N)| \;\sim\;
    \begin{cases}
        N^{-0.284} & T = \infty, \\[4pt]
        N^{-0.245} & T = 0.1,
    \end{cases}
    \label{eq:peak_separation_p}
\end{equation}
more slowly than the finite-size critical point $p_c(N)$ itself does to the true critical point, according to $1/\bar{\nu}$ in \Cref{tab:exponents}. Consequently, the peak enters a regime in which the rescaled deviation $(p_{{peak}(N)} - p_c(N))\,N^{1/\bar\nu}$ diverges with $N$, meaning the peak does not necessarily follow the critical scaling function in finite systems. Instead, it develops its own size-dependent scaling that is controlled by the routing ensemble rather than by the structural universality class. 

The temperature dependence of $l_{\max}$ reveals how the routing ensemble shapes this precursor. At low temperature, $l_{\max} \sim N^{1/3}$, indicating that the routing process has exhausted geodesic paths up to the mean-field chemical correlation length imposed by the crossover horizon. At high temperature, noisy routing explores longer paths, and $l_{\max}$ increases but grows more slowly than $N^{1/3}$, which also implies that at $p_{peak}$ or $t_{peak}$, the giant component is far from a tree-like structure typical of mean-field criticality. Crucially, despite this difference in the pre-critical peak, the critical path length $l_c$ scales similarly in both temperature regimes, confirming that the structural universality class at $t_c$ is insensitive to the routing temperature.

\begin{figure}
    \centering
    \includegraphics[width=\linewidth]{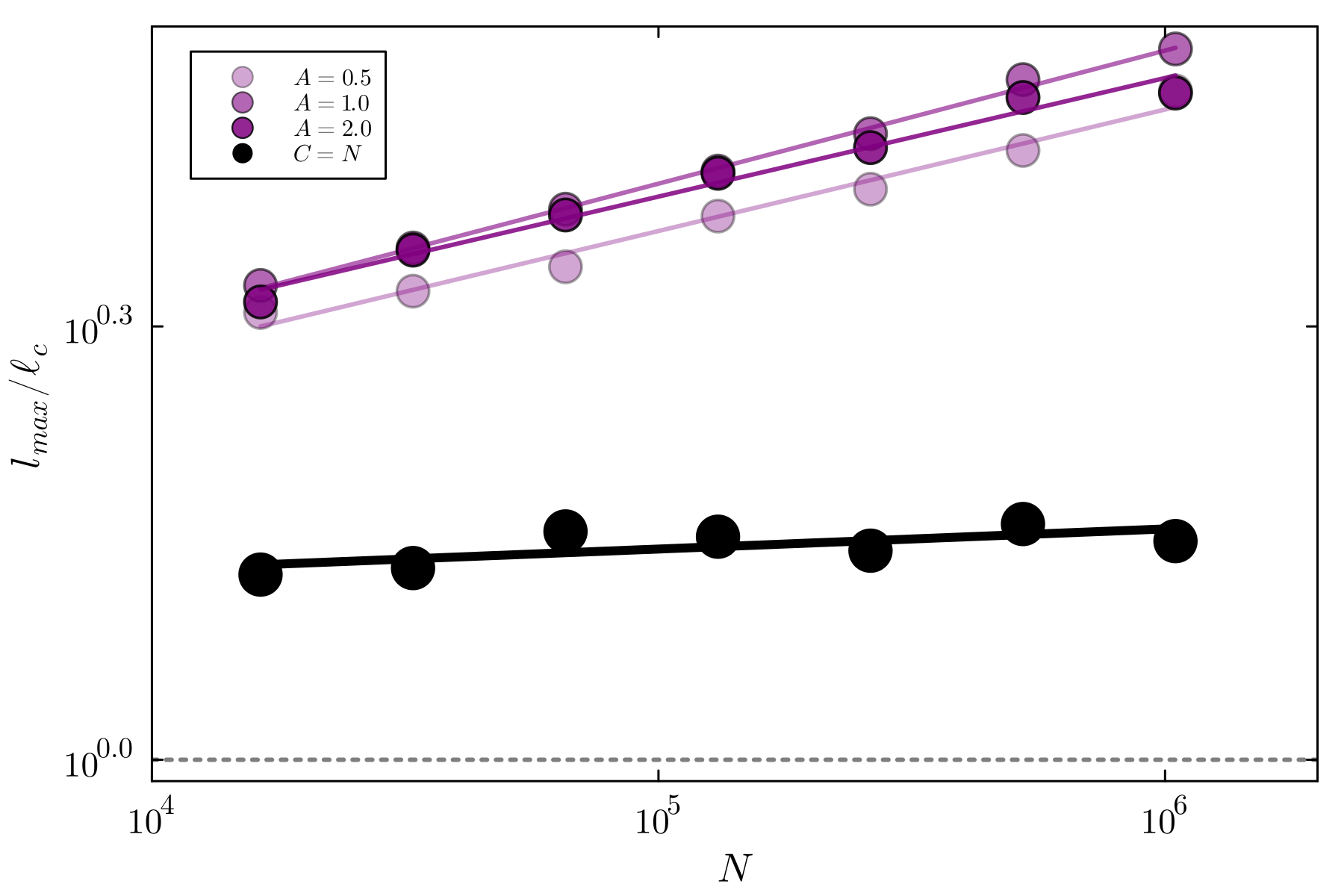}
    \caption{Ratio $l_{\max}/l_c$ between the pre-critical path-length peak and the critical path length, as a function of system size $N$, for different horizon prefactors $A$ in $C = A\,N^{1/3}$ (source-uniform ensemble, $T = 0.1$). For the finite crossover horizons ($A = 0.5, 1, 2$) the ratio grows steadily with $N$. For the uncapped horizon $C = N$ (black), the ratio is nearly $N$-independent.}
    \label{fig:ratio}
\end{figure}

Together, these results establish that in the crossover regime $C = N^{1/3}$, the path ensemble encodes a dynamical precursor to network failure that is decoupled from the structural critical point. The precursor is established at the level of timescales: the path-length peak escapes the critical time window, with $|t_c(N) - t_{peak}(N)|$ diverging faster than the window width $N^{\,1-a_c-1/\bar\nu}$, so that path elongation and giant-component collapse become asymptotically distinct events. The length-scale signature, $l_{\max}$, reflects this separation, with a magnitude set by the routing details, i.e., the horizon and temperature, rather than by the percolation universality class alone. The same precursor persists across other moderate prefactors, including $C = 0.5\,N^{1/3}$ and $C = 2\,N^{1/3}$. For the uncapped horizon $C = N$, the precursor's length-scale signature weakens: the scaling exponents of $l_{\max}$ and $l_c$ drift toward each other and the ratio $l_{\max}/l_c$ grows only marginally with $N$, in contrast to the finite crossover horizons where it increases more steadily (\Cref{fig:ratio}). This indicates the recoupling of path elongation and structural fragmentation as the horizon saturates.


\begin{table}[t]
\centering
\caption{Critical exponents from finite-size scaling on Newman--Watts networks. All finite-$C$ cases
are consistent with mean-field values ($1/\bar{\nu} = \beta/\bar{\nu} = \gamma/\bar{\nu} = 1/3$). Growing horizons
$C = A\,N^{1/3}$ drive the system into a distinct crossover regime, with exponent estimates that tend towards the saturated-horizon ($C=N$) values as the prefactor $A$ increases. System sizes, the number of simulations, and example log-log plots are given in the Supplementary Materials.}
\label{tab:exponents}
\renewcommand{\arraystretch}{1.2}
\begin{tabular}{@{}llccc@{}}
\toprule
Ensemble & \(C\) & \(1/\bar{\nu}\) & \(\beta/\bar{\nu}\) & \(\gamma/\bar{\nu}\) \\
\midrule

\multicolumn{5}{@{}l@{}}{\textit{Pair-uniform}} \\
\quad $T=0.1$    & $3$ & $0.343\pm0.003$ & $0.323\pm0.004$ & $0.342\pm0.002$ \\
\quad $T=\infty$ & $3$ & $0.339\pm0.001$ & $0.332\pm0.004$ & $0.327\pm0.004$ \\[3pt]

\multicolumn{5}{@{}l@{}}{\textit{Source-uniform}} \\
\quad $T=0.1$    & $3$ & $0.333\pm0.004$ & $0.326\pm0.003$ & $0.330\pm0.002$ \\
\quad $T=\infty$ & $3$ & $0.333\pm0.002$ & $0.323\pm0.003$ & $0.335\pm0.004$ \\[3pt]
\quad $T=0.1$    & $0.5N^{1/3}$ & $0.388\pm0.005$ & $0.254\pm0.001$ & $0.446\pm0.003$ \\
\quad $T=0.1$    & $N^{1/3}$ & $0.423\pm0.004$ & $0.248\pm0.001$ & $0.468\pm0.003$ \\
\quad $T=\infty$ & $N^{1/3}$ & $0.453\pm0.004$ & $0.240\pm0.001$ & $0.492\pm0.003$ \\
\quad $T=0.1$    & $2N^{1/3}$ & $0.434\pm0.006$ & $0.233\pm0.001$ & $0.484\pm0.002$ \\
\quad $T=0.1$ & $N$ & $0.491 \pm 0.004$ & $0.211 \pm 0.002$ & $0.527 \pm 0.002$ \\
\quad $T=\infty$ & $N$ & $0.498 \pm 0.002$  & $0.212 \pm 0.001$ & $0.535 \pm 0.001$\\
SPP (Ref.~\cite{kim_shortest-path_2024}) & $N$ & $ - $ & $0.21$ & $0.55$ \\
\midrule
Mean-field & --- & $1/3$ & $1/3$ & $1/3$ \\
\bottomrule
\end{tabular}
\end{table}

\clearpage
\newpage    
\section{Conclusion and discussion}
We have introduced a generalised path-percolation framework that extends shortest-path percolation by disentangling three ingredients that are coupled in the original model: the origin--destination ensemble, the routing protocol, and the routing horizon. This separation allows us to distinguish the features of flow-induced damage that control universal critical behaviour from those that primarily affect non-universal quantities such as the percolation threshold.

For finite routing horizons, path-induced correlations remain short-ranged and can therefore be coarse-grained out by box covering. Consistent with this picture, the measured critical exponents remain close to their mean-field values despite substantial differences in the microscopic removal dynamics. The principal effect of routing at finite C is instead on non-universal quantities, most notably the critical threshold $p_c$.

We showed that shifts in $p_c$ are closely linked to the localisation of the edge-load distribution. At low routing temperature, traffic is concentrated on a small subset of high-load edges, generating an effectively targeted damage process and accelerating fragmentation. At high temperature, the load is distributed over a larger fraction of the network, producing more homogeneous damage and increasing $p_c$. The entropy and participation ratio of the load distribution therefore provide natural microscopic observables connecting routing organisation to macroscopic robustness.

The origin--destination ensemble is likewise an active component of the dynamics rather than a passive modelling choice. Pair-uniform sampling probes globally selected connected pairs and therefore emphasises paths that support large-scale connectivity. Source-uniform sampling, by contrast, represents demand initiated from local seeds. In this ensemble, flow can be accommodated inside finite connected components, so local demand may remain satisfiable even when it does not contribute to the giant component. For finite $C$, the distinction modifies the percolation point while leaving the critical exponents consistent with mean-field behaviour. 

The most pronounced departure from finite-$C$ behaviour occurs when the routing horizon grows as $C=N^{1/3}$, matching the mean-field chemical correlation length. In this regime, path-induced correlations are no longer finite-range perturbations, and source-uniform sampling allows finite clusters near criticality to participate directly in the flow process. Together, these effects produce exponent estimates distinct from both finite-$C$ mean-field behaviour and path percolation with an infinite budget.

A striking consequence of this crossover is the dynamical decoupling of path elongation from structural fragmentation. The characteristic path length develops a growing peak, shaped by the routing ensemble, strictly before the collapse of the giant component: the separation $|t_c - t_{peak}|$ diverges faster than the width of the critical time window, so the two events become asymptotically distinct. Meanwhile, the critical path length $l_c$ follows the universal scaling and is insensitive to the routing temperature. Flow-induced failure can thus exhibit a dynamical precursor encoded in the path ensemble itself.

Taken together, our results show that microscopic routing organisation and demand dispatch jointly determine network robustness. More broadly, path ensembles play a role analogous to interaction rules in correlated percolation, with the range of path-induced correlations providing a natural organising principle for universality and crossover behaviour in flow-driven fragmentation.

Several questions remain open. Most importantly, an analytic theory of the crossover regime is needed to determine whether the exponent estimates obtained at $C = A\,N^{1/3}$ correspond to genuinely ensemble-dependent asymptotic behaviour or to a slow crossover towards the saturated-horizon limit. The monotonic drift of the estimates with the prefactor $A$, together with the weakening of the precursor as the horizon saturates towards $C = N$, suggests that the scaling of $C$ with system size acts as a relevant perturbation, which remains to be characterised. Future work should also test the generality of the pre-critical path-length precursor on other network substrates, and explore routing strategies on weighted and multiplex networks that deliberately delocalise load to enhance robustness.

\section*{Acknowledgements}
The authors acknowledge the use of the University of Oxford Advanced Research Computing (ARC) facility in carrying out this work (\href{https://doi.org/10.5281/zenodo.22558}{https://doi.org/10.5281/zenodo.22558}). Y.D. acknowledges financial support from the CSC--PAG Scholarship funded by Oriel College and the China Scholarship Council.

\section*{Data Availability}

The data and source code that support the findings of this study are available from the corresponding author upon reasonable request.

\printbibliography

\clearpage

\begin{center}
{\LARGE\bfseries Supplementary Material}
\end{center}
\renewcommand{\thesection}{S\arabic{section}}
\renewcommand{\theHsection}{S\arabic{section}}
\setcounter{section}{0}
\renewcommand{\thefigure}{S\arabic{figure}}
\renewcommand{\theHfigure}{S\arabic{figure}}
\setcounter{figure}{0}
\renewcommand{\thetable}{S\arabic{table}}
\renewcommand{\theHtable}{S\arabic{table}}
\setcounter{table}{0}
\renewcommand{\theequation}{S\arabic{equation}}
\renewcommand{\theHequation}{S\arabic{equation}}
\setcounter{equation}{0}
\renewcommand{\thealgorithm}{S\arabic{algorithm}}
\renewcommand{\theHalgorithm}{S\arabic{algorithm}}
\setcounter{algorithm}{0}
\input{Supplmentary_material}
\end{document}

%% file: Supplmentary_material.tex
\section{Simulation Procedure}

This section describes the numerical implementation of the temperature-based path percolation model. Each simulation generates a complete edge-removal sequence, which is subsequently converted into a bond-addition process for computing percolation observables. The overall procedure consists of three stages: (i) stochastic sampling of an origin--destination pair, (ii) temperature-dependent path construction, and (iii) edge removal along the sampled path. The process is repeated until all edges have been removed.

For pair-uniform demand, we refer the readers to the work \cite{kim_shortest-path_2024}. For source-uniform ensemble, we have the following protocol. At each iteration, an origin and a destination are selected uniformly at random from the set of active nodes, i.e., nodes with at least one remaining incident edge. A breadth-first search (BFS) is then performed from the origin to construct a local search tree with maximum depth $C$. If the destination is unreachable within the truncated BFS tree, it is replaced by a uniformly selected reachable node. Marginalising over the initial draw of the destination $d$, the probability that a given node $x \in  B^{(C)}_t(o)$ is selected as the destination is
\begin{equation}
\mathbb{P}(d = x)
= \underbrace{\frac{1}{|V_t|-1}}_{\text{direct hit}}
+ \underbrace{\left(1 - \frac{| B^{(C)}_t(o)|}{|V_t|-1}\right)
  \frac{1}{| B^{(C)}_t(o)|}}_{\text{miss, then redraw}}
= \frac{1}{| B^{(C)}_t(o)|},
\label{eq:replacement-exact}
\end{equation}
where the first term corresponds to the initial draw landing on $x$ directly,
and the second to the initial draw falling outside $B$ (probability
$1-|B|/(|V_t|-1)$) followed by a uniform redraw within the ball. Since the
result is uniform on $B^{(C)}_t(o)$ for every origin, the sampled pairs are
distributed exactly according to the source-uniform ensemble $Q^2_t(o,d)$, and the replacement procedure serves only to avoid rejection resampling.

A stochastic path is subsequently generated using the preferential walk described in \Cref{alg:walk}. The transition probability depends on the routing temperature $T_x$, the degree preference parameter $\alpha$, and the remaining graph structure. Finally, each edge on the sampled path is removed independently according to the removal temperature $T_s$. The procedure terminates once every edge has been removed. 

The full history of removed edges defines an ordered edge-removal sequence. This sequence is reversed and interpreted as a bond-addition process. Percolation observables are computed using a Union--Find algorithm, which efficiently tracks connected components during reconstruction. The number of realisations for different system size is summarised in \Cref{tab:simulation_numbers}.
\begin{table}[ht!]
\centering
\caption{Number of independent simulation realisations for each system size.}
\begin{tabular}{cc}
\toprule
System sizes $N$ & Simulations \\
\midrule
$2^{14}$--$2^{20}$ & 4800 \\
$2^{21}$ & 1600 -- 4800 \\
$2^{22}$ & 1600 -- 4800 \\
$2^{23}$ & 500\\
\bottomrule
\end{tabular}
\label{tab:simulation_numbers}
\end{table}

end whilereturn\begin{algorithm}[H]
\caption{Preferential Walk: sample path $\pi$ from $d$ to $o$}
\label{alg:walk}
\begin{algorithmic}[1]
\Require BFS distances $\{l_v\}$ from $o$; parameters $\alpha,\Tx$;
         horizon $C$; current degrees $\{k_v\}$
\Ensure  Path $\pi=[d,\ldots,o]$ and corresponding edge list, or \textsc{Fail}
\State $\text{cur} \leftarrow d$;\quad
       $\pi \leftarrow [d]$;\quad
       $\mathcal{E} \leftarrow []$;\quad
       $\text{visited} \leftarrow \{d\}$
\While{$\text{cur} \neq o$}
  \State $\mathcal{N} \leftarrow \varnothing$
  \For{each unremoved neighbour $(v, e)$ of $\text{cur}$}
    \If{$l_v = \infty$ \textbf{or} $v \in \text{visited}$}
      \textbf{continue}
    \EndIf
    \If{$l_v \leq l_{\text{cur}}$} \algorithmiccomment{distance-descent constraint}
      \State $\Delta l \leftarrow l_{\text{cur}} - l_v$
      \State $w(v) \leftarrow
             \bigl(k_{\text{cur}}\,k_{v}\bigr)^{\alpha}
             \exp\!\left(\dfrac{\Delta l}{\Tx}\right)$
      \State $\mathcal{N} \leftarrow \mathcal{N} \cup \{(v,e,w(v))\}$
    \EndIf
  \EndFor
  \If{$\mathcal{N} = \varnothing$}
    \Return \textsc{Fail}
  \EndIf
  \State Draw $(v^*,e^*) \sim \mathcal{N}$ with probability $\propto w(v)$
  \State $\pi \mathrel{+}= v^*$;\quad
         $\mathcal{E} \mathrel{+}= e^*$;\quad
         $\text{visited} \leftarrow \text{visited} \cup \{v^*\}$
  \State $\text{cur} \leftarrow v^*$
\EndWhile
\State \Return $\pi,\, \mathcal{E}$
\end{algorithmic}
\end{algorithm}
 
\vspace{0.3cm}

\begin{algorithm}[H]
\caption{Temperature-Based Stochastic Path Percolation in Source-Uniform Ensemble}
\label{alg:main}
\begin{algorithmic}[1]
\Require Graph $G=(V,E)$;
         disorder fields $\{x_e\}$, $\{s_e\}$ on edges;
         parameters $\Tx,\Ts,\alpha$; BFS horizon parameter $C$
\Ensure  Ordered edge-removal sequence $\mathcal{L}$
 
\State \textbf{Initialise:}\quad
       $k_u \leftarrow \deg(u)\ \forall u$;\quad
       $\text{removed}[e] \leftarrow \textsc{False}\ \forall e$
\State $\Adeg \leftarrow \{u \in V : k_u > 0\}$
       \algorithmiccomment{active-node pool}
\State $R \leftarrow |E|$;\quad $\mathcal{L} \leftarrow []$
 
\While{$R > 0$}
 
  \vspace{2pt}
  \State \mycomment{\textit{--- Step 1: sample O-D pair}}
  \State Draw $o, d$ uniformly at random from $\Adeg$ (distinct)
 
  \vspace{2pt}
  \State \mycomment{\textit{--- Step 2: BFS from origin ---}}
  \State Run BFS from $o$ on remaining edges, cutting at depth $C$
  \If{$d$ is unreachable}
  
    \State $d \leftarrow $ uniformly random node in $\text{BFS-tree}(o)$
  \EndIf
 
  \vspace{2pt}
  \State \mycomment{\textit{--- Step 3: preferential walk ---}}
  \State $(\pi,\mathcal{E}) \leftarrow \textsc{PreferentialWalk}(o,d,\Tx,\alpha)$
         \algorithmiccomment{Algorithm~\ref{alg:walk}}
  \If{$\pi = \textsc{Fail}$}
    \textbf{continue}
  \EndIf
 
  \vspace{2pt}
  \State \mycomment{\textit{--- Step 4: stochastic edge removal ---}}
  \For{each edge $e \in \mathcal{E}$ with endpoints $(u,v)$}
    \If{$\text{removed}[e]$} \textbf{continue} \EndIf
    \State Remove $e$ with probability $\exp\!\left(-s_e / \Ts\right)$
    \If{$e$ removed}
      \State $\mathcal{L} \mathrel{+}= e$;\quad
             $k_u \mathrel{-}= 1$;\quad $k_v \mathrel{-}= 1$;\quad
             $R \mathrel{-}= 1$
      \If{$k_u = 0$} remove $u$ from $\Adeg$ \EndIf
      \If{$k_v = 0$} remove $v$ from $\Adeg$ \EndIf
    \EndIf
  \EndFor
 
\EndWhile

\Return $\mathcal{L}$
\end{algorithmic}
\end{algorithm}
 
\vspace{0.3cm}

\renewcommand{\arraystretch}{1.3}
\begin{tabular}{c l}
\toprule
Parameter & Interpretation \\
\midrule
$\Tx \to 0$      & Geodesic (shortest-path) limit \\
$\Tx \to \infty$  & Diffusive/random walk limit \\
$\Ts \to 0$      & No removal (flow process) \\
$\Ts \to \infty$ & Deterministic removal\\
$\alpha > 0$     & Hub-favoring flows \\
$\alpha < 0$     & Hub-avoiding flows \\
\bottomrule
\end{tabular}

\vspace{4mm}

In this work, we consider the model with $T_x = T$, $T_s = \infty$, $s_e = 1$, and $\alpha = 0$. More generally, the framework allows heterogeneous edge disorder through edge-dependent removal fields $s_e$ and finite removal temperature $T_s$, corresponding to soft (probabilistic) dismantling rather than deterministic edge removal. These ingredients provide a more general description of path-induced network degradation, although they are not explored in the present work.

\begin{figure}[ht!]
    \centering
    \includegraphics[width=1\linewidth]{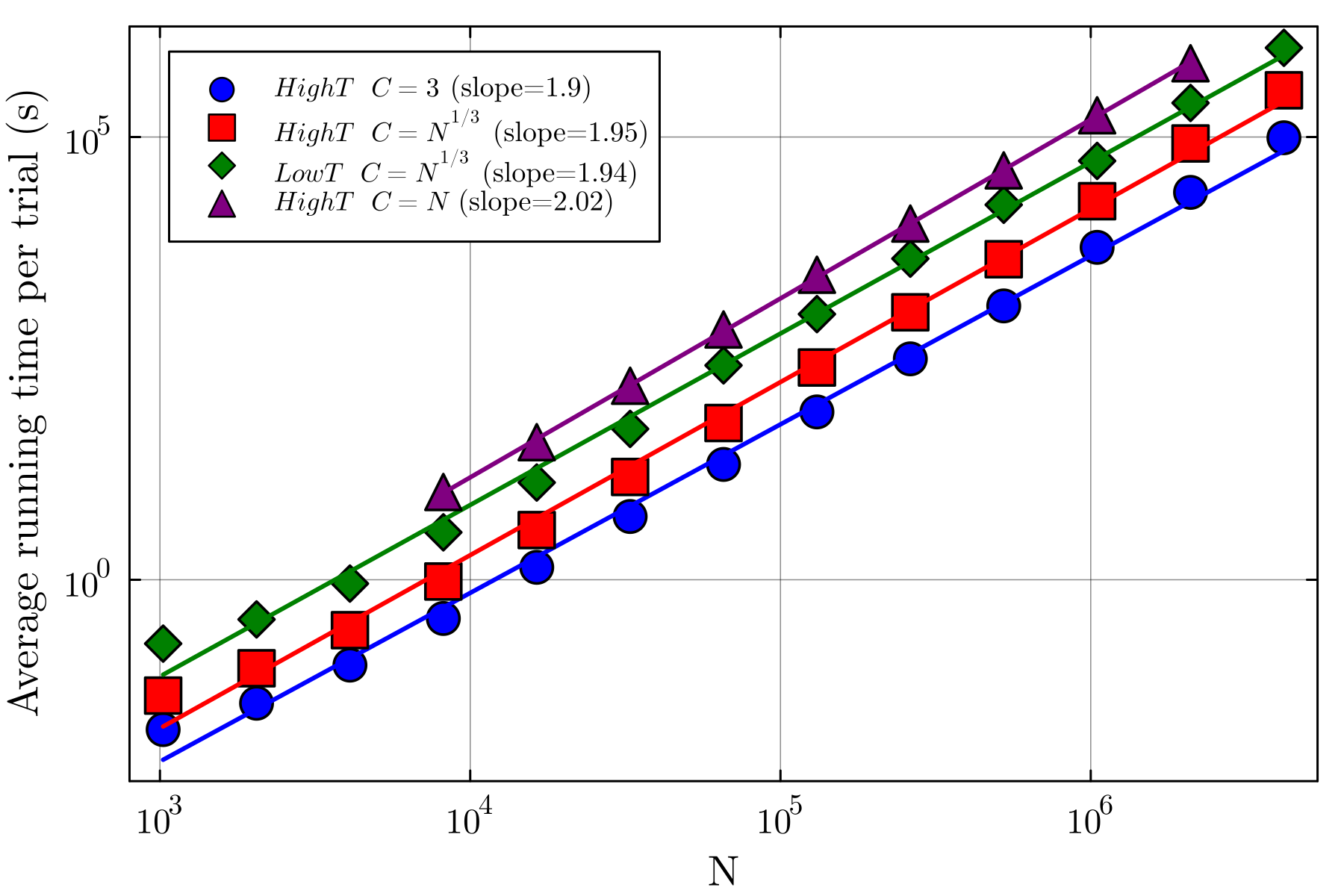}\caption{Empirical runtime of a single simulation as a function of system size for representative parameter values in the source-uniform ensemble. The observed scaling is approximately $T(N)\sim N^{1.9\text{--}2.0}$}
    \label{fig:timecomplexity}
\end{figure}

The computational cost is dominated by repeated truncated BFS searches and stochastic path sampling. Empirically, the runtime of a single realisation scales approximately as
$T(N)\sim N^{1.9\text{--}2.0}$, over the range of system sizes and parameter values considered in the source-uniform ensemble. The measured runtime is shown in \Cref{fig:timecomplexity}.

\section{The shortest-path region of the routing temperature}
\label{sec:sp-region}

In the main text, low-temperature rows are identified with shortest-path
routing (``$T \to 0$''). Here we quantify this identification by estimating
the temperature range within which the temperature-controlled walk is
equivalent to geodesic routing for all observables considered in this work.

The routing protocol constrains the walker to non-ascending moves with
weight $w = \exp(\Delta l/T)$, where $\Delta l \in \{0, 1\}$ is the decrease
in BFS distance to the origin. Suppose that at the current node $c$ the
walker has $n_1$ descending ($\Delta l = 1$) and $n_0$ horizontal
($\Delta l = 0$) options. The probability of taking a horizontal step is
\begin{equation}
p_{\mathrm{lat}}
= \frac{n_0}{\,n_1\, e^{1/T} + n_0\,}
\;\approx\; r\, e^{-1/T},
\qquad r \equiv \frac{n_0}{n_1},
\label{eq:plat}
\end{equation}
where the approximation holds for $e^{1/T} \gg 1$. The factor $e^{-1/T}$ is
the Boltzmann weight of a horizontal move on the BFS energy landscape, whose
energy gap relative to descent is unity; the ratio $r$ is a prefactor set by
the substrate. In the Newman--Watts model with $K = 4$ and $\beta = 0.1$, we
estimate $r$ to lie between $0.5$ and $2$ on average.

We define the shortest-path region by requiring that the expected number of
horizontal moves per sampled path, $r\,l\, e^{-1/T}$, remains below unity
for all sampled path lengths. The largest sampled length in the main text is
$l_{\max} \approx 120$ at $N = 2^{22}$, which yields the upper boundary
\begin{equation}
T^{*} = \frac{1}{\ln\!\left(r\,l_{\max}\right)} \in [0.18,\, 0.24].
\label{eq:Tstar}
\end{equation}
Meanwhile, the statistical error across realisations scales as $n^{-1/2}$,
which for $n = 1600$--$4800$ trials amounts to $1.4$--$2.5\%$. Taking
$T = 0.2$ as an example, the relative excess in path length is
$r\, e^{-1/T} = r\, e^{-5} \approx 0.35$--$1.4\%$, well below the
statistical error. This also shows that in the low-temperature region,
horizontal moves contribute only an $N$-independent prefactor to the sampled path length, and therefore cannot modify any of the scaling exponents reported in the main text.

\section{Finite-Size Scaling Analysis}

The critical exponents are estimated using finite-size scaling in the event-based ensemble \cite{fan_universal_2020}. For a finite system of size $N$, the pseudo-critical occupation probability $p_c(N)$ converges to the thermodynamic critical point according to
\begin{equation}\label{eq:ffs_pc}
p_c(N)-p_c\sim N^{-1/\bar{\nu}}.
\end{equation}

Near the critical point, the order parameter and susceptibility satisfy the finite-size scaling forms
\begin{align}
S(p,N)
&=
N^{-\beta/\bar{\nu}}
\Phi_S\!\left[(p-p_c)N^{1/\bar{\nu}}\right],
\label{eq:sc}
\\
\chi(p,N)
&=
N^{\gamma/\bar{\nu}}
\Phi_\chi\!\left[(p-p_c)N^{1/\bar{\nu}}\right].
\label{eq:chi}
\end{align}

For each realisation, the pseudo-critical point $p_c(N)$ is identified as the bond occupation fraction at which the order parameter exhibits its largest jump (we have verified using the peak of $\chi$ does not result in a qualitative difference in estimates). The corresponding order parameter and susceptibility satisfy
\begin{align}\label{eq:S}
\langle S(p_c(N),N)\rangle
&\sim
N^{-\beta/\bar{\nu}},
\\
\langle\chi(p_c(N),N)\rangle
&\sim
N^{\gamma/\bar{\nu}}.
\end{align}

The exponent ratios $\beta/\bar{\nu}$ and $\gamma/\bar{\nu}$ are obtained by linear regression on logarithmic scales. The reported uncertainties correspond to the standard errors of the fitted slopes and therefore quantify the statistical uncertainty of the regression rather than systematic finite-size corrections.

The asymptotic critical point $p_c$ is determined by scanning candidate values in the conventional ensemble. Candidate values are sampled around the average pseudo-critical point $p_c(N_{\max})$ using the data $S(p,N)$. For each candidate, we fit the scaling relation $S(p_c,N)\sim N^{-\beta^{*}/\bar{\nu}^{*}}.$ The optimal estimate is selected as the value that minimises the difference between the fitted exponent ratio $\beta^{*}/\bar{\nu}^{*}$ and the independently estimated value $\beta/\bar{\nu}$ obtained from \Cref{eq:S}, while maintaining a high coefficient of determination ($R^2$), indicating a good power-law fit. The exponent $\bar{\nu}$ is then estimated either from the finite-size shift relation, \Cref{eq:ffs_pc}, or from the event-based scaling relation
$\sigma\!\left(p_c(N)\right)\sim N^{-1/\bar{\nu}}.$ Both approaches yield consistent estimates.

\begin{figure}
    \centering
    \includegraphics[width=1\linewidth]{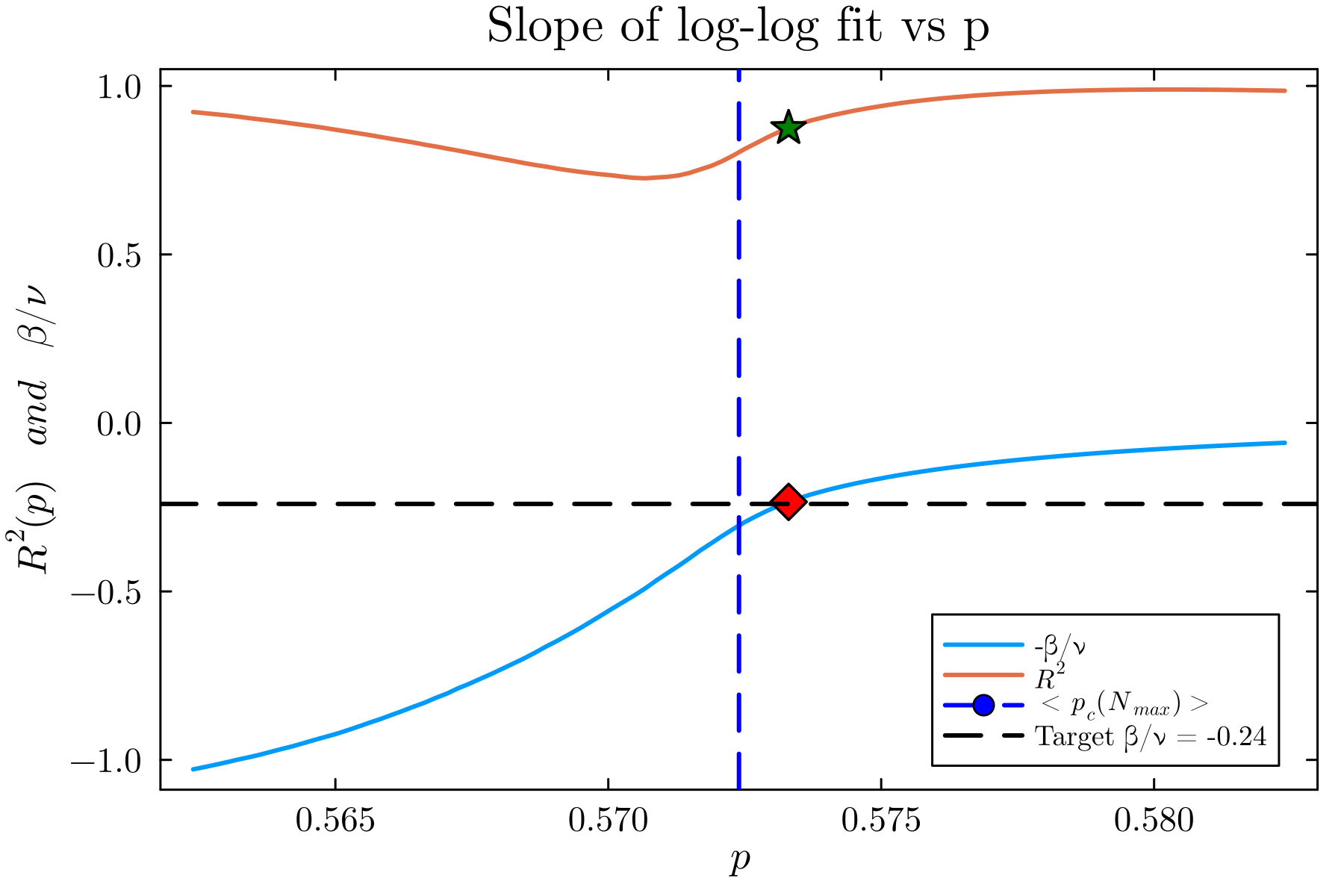}
    \caption{The error analysis for estimating $p_c$ in source-uniform ensemble with $C = N^{1/3}$ and $T = \infty$. The green star and red diamond denote the $\beta/\bar{\nu}$ and the $R^2$ at $p_c$. The targeted slope is estimated from \Cref{eq:S}.}
    \label{fig:pc_error_analysis}
\end{figure}

\begin{algorithm}[H]
\caption{Percolation Diagram and Finite Size Scaling}
\label{alg:fss}
\begin{algorithmic}[1]
\Require Removal sequence $\mathcal{L}$ (per trial); number of trials $\mathcal{T}$
\Ensure  Critical exponents $\beta/\bar{\nu},\,\gamma/\bar{\nu},\,1/\bar{\nu}$ and $p_c$
 
\For{trial $t = 1,\ldots,\mathcal{T}$}
  \State Run Algorithm~\ref{alg:main} $\Rightarrow$ removal sequence $\mathcal{L}^{(t)}$
  \State \textbf{Reverse} $\mathcal{L}^{(t)}$ and rebuild $G$ edge by edge (bond-addition order)
  \For{each bond-occupation fraction $p = k/|E|$, $k=1,\ldots,|E|$}
    \State $\GCC(p) \leftarrow $ largest component size $/ N$
    \State $\chi(p) \leftarrow \sum_{i \neq \text{GCC}} s_i^2 \,/\, N$
           \algorithmiccomment{susceptibility}
  \EndFor
  \State $p_c^{(t)} \leftarrow \operatorname*{arg\,max}_{p_k}\left[S^{(t)}(p_{k+1})-S^{(t)}(p_k)\right] $; \quad
         $S_c^{(t)} \leftarrow \GCC\!\left(p_c^{(t)}\right)$;\quad
         $\chi_c^{(t)} \leftarrow \chi\!\left(p_c^{(t)}\right)$
\EndFor
 
\State Average over trials:
       $\langle p_c \rangle,\;\langle S_c \rangle,\;\langle \chi_c \rangle$
 
\State \mycomment{\textit{--- Finite-size scaling fits (log-log regression) ---}}
\State Fit $\langle S_c(N)\rangle \sim N^{-\beta/\bar{\nu}}$
\State Fit $\langle \chi_c(N)\rangle \sim N^{\,\gamma/\bar{\nu}}$
\State Fit $\sigma(p_c(N)) \sim N^{-1/\bar{\nu}}$
\State Scan candidate p within $(\langle p_c\rangle  -\Delta, \langle p_c \rangle+ \Delta)$, calculate the slope and the goodness of fit $R^2$, $p_c = argmin_{p}(|\beta^{*}/\bar{\nu}^{*} - \beta/\bar{\nu}|)$. 
\State Fit $|\langle p_c(N)\rangle - p_c^\infty| \sim N^{-1/\bar{\nu}}$ $\Rightarrow$ exponent $1/\bar{\nu}$
\State \Return $p_c,\;\beta/\bar{\nu},\;\gamma/\bar{\nu},\;1/\bar{\nu}$
\end{algorithmic}
\end{algorithm}

\vspace{0.3cm}

\begin{figure}[!ht]
    \centering

    \begin{subfigure}{\textwidth}
        \centering
        \includegraphics[width=\linewidth]{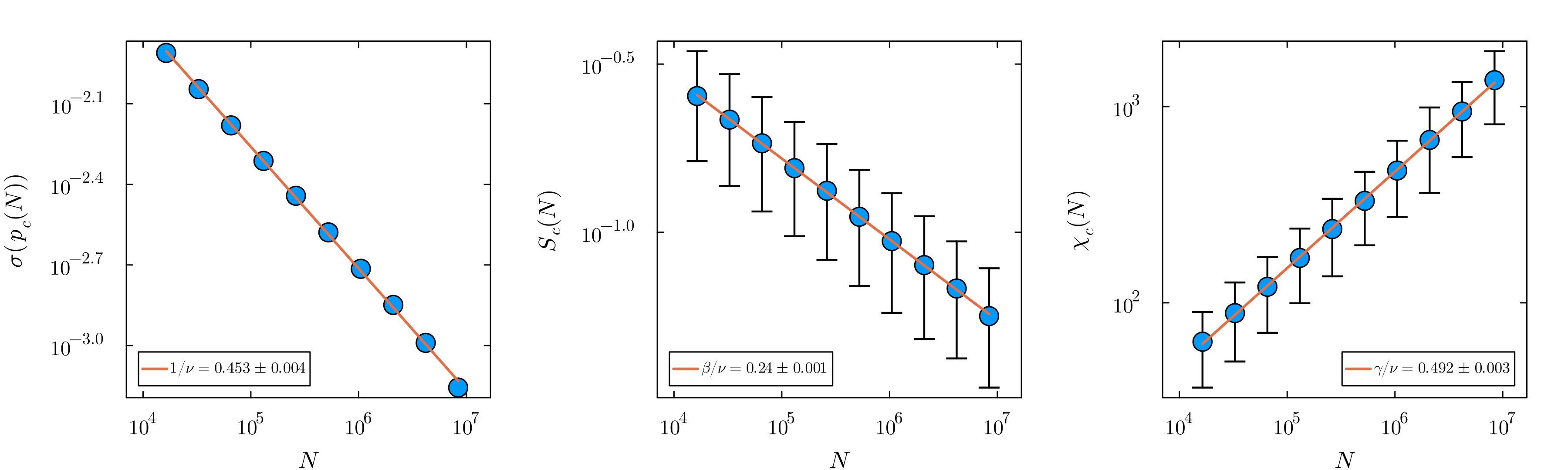}
        \caption{Critical exponents for path percolation with $T=\infty$ and $C=N^{1/3}$.}
        \label{fig:CriticalExponents_crossover_source_uni_high_T}
    \end{subfigure}

    \vspace{0.5em}

    \begin{subfigure}{\textwidth}
        \centering
        \includegraphics[width=\linewidth]{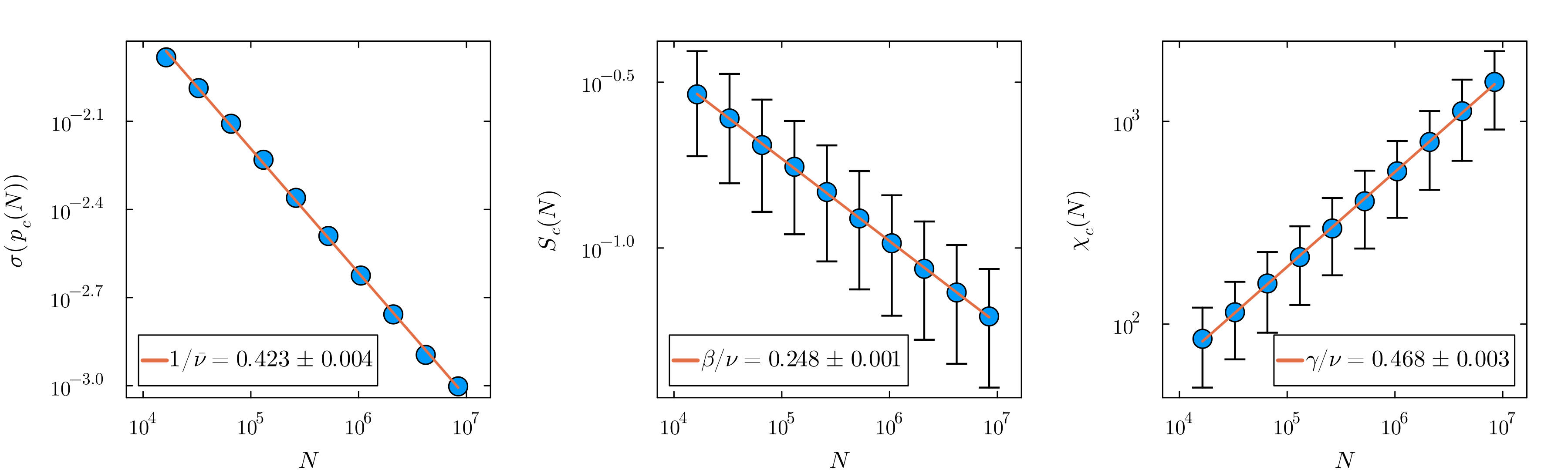}
        \caption{Critical exponents for path percolation with $T=0.1$ and $C=N^{1/3}$.}
        \label{fig:CriticalExponents_crossover_source_uni_low_T}
    \end{subfigure}

    \caption{
    Finite-size scaling analysis for the source-uniform ensemble with a growing routing horizon $C=N^{1/3}$ in the Newman--Watts model.
    }
    \label{fig:CriticalExponents_crossover_source_uni}
\end{figure}

\section{Simulation on the heterogeneous networks}
We also generate the clustered-heterogeneous networks using the Holme-Kim model \cite{holme_growing_2002} and perform the simulation with different routing temperatures in a pair-uniform ensemble.

\begin{figure}[!ht]
    \centering
    \includegraphics[width=1\linewidth]{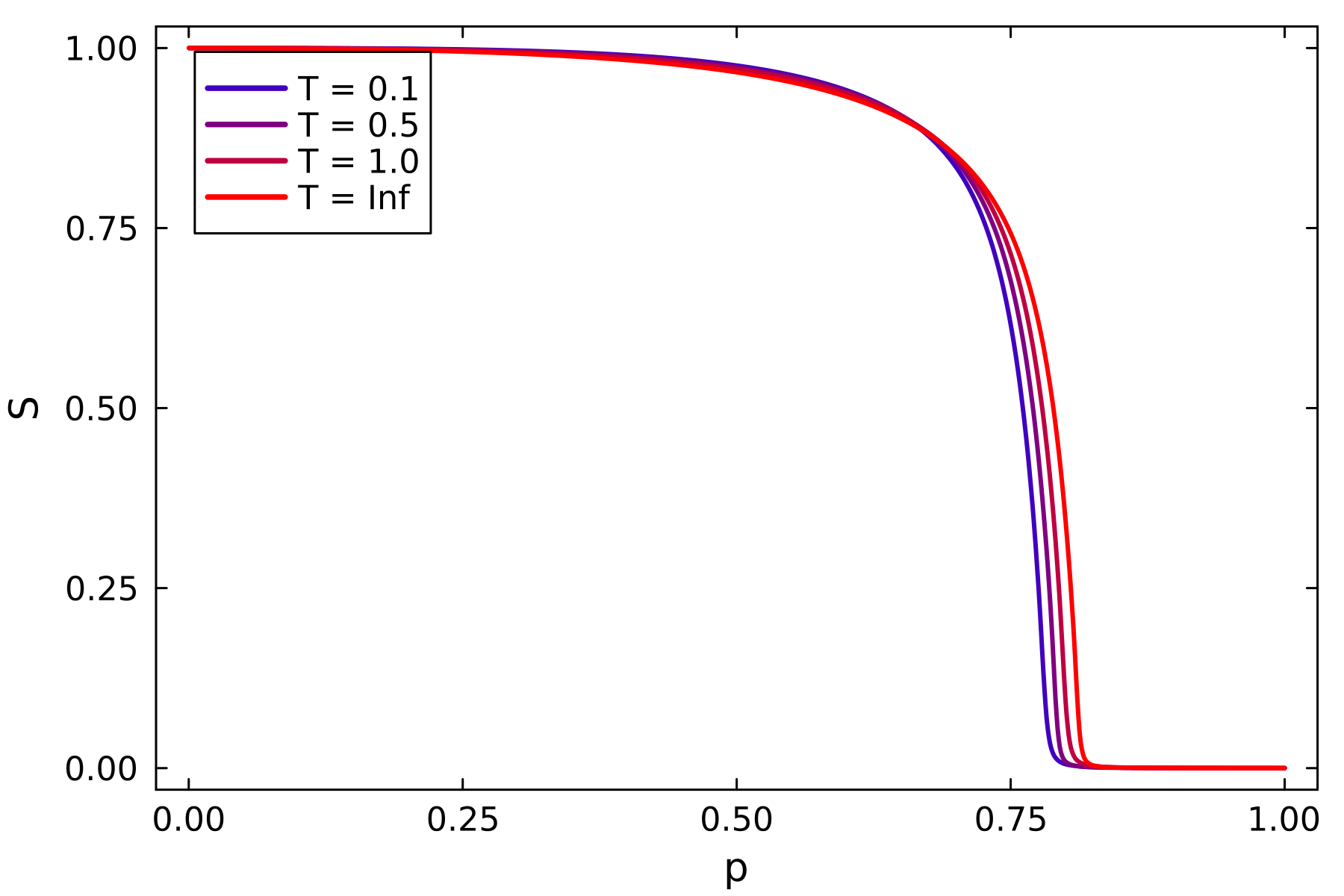}
    \caption{Path percolation with different routing temperatures on Holme-Kim networks with $2^{19}$ nodes and the triad formation probability $0.9$.}
    \label{fig:Holme-Kim}
\end{figure}

\section{Participation Ratio as a Control Parameter for Network Robustness}

The normalised participation ratio (PR) provides an alternative measure of traffic delocalisation to the entropy defined in the main text. Let $p_{ij}$ denote the normalized traffic carried by the edge connecting nodes $i$ and $j$, satisfying
\[
\sum_{(i,j)\in E} p_{ij}=1,
\]
where $E$ is the set of edges. The normalised participation ratio is defined as
\begin{equation}
\mathrm{PR}
=
\frac{1}{|E|\displaystyle\sum_{(i,j)\in E} p_{ij}^{\,2}},
\end{equation}
where $|E|$ is the total number of edges. The normalised participation ratio satisfies $1/|E| \leq \mathrm{PR} \leq 1$, with larger values indicating a more homogeneous distribution of traffic over the network.
\begin{figure}[!ht]

    \centering

    \begin{subfigure}{\textwidth}

        \centering

        \includegraphics[width=\linewidth]{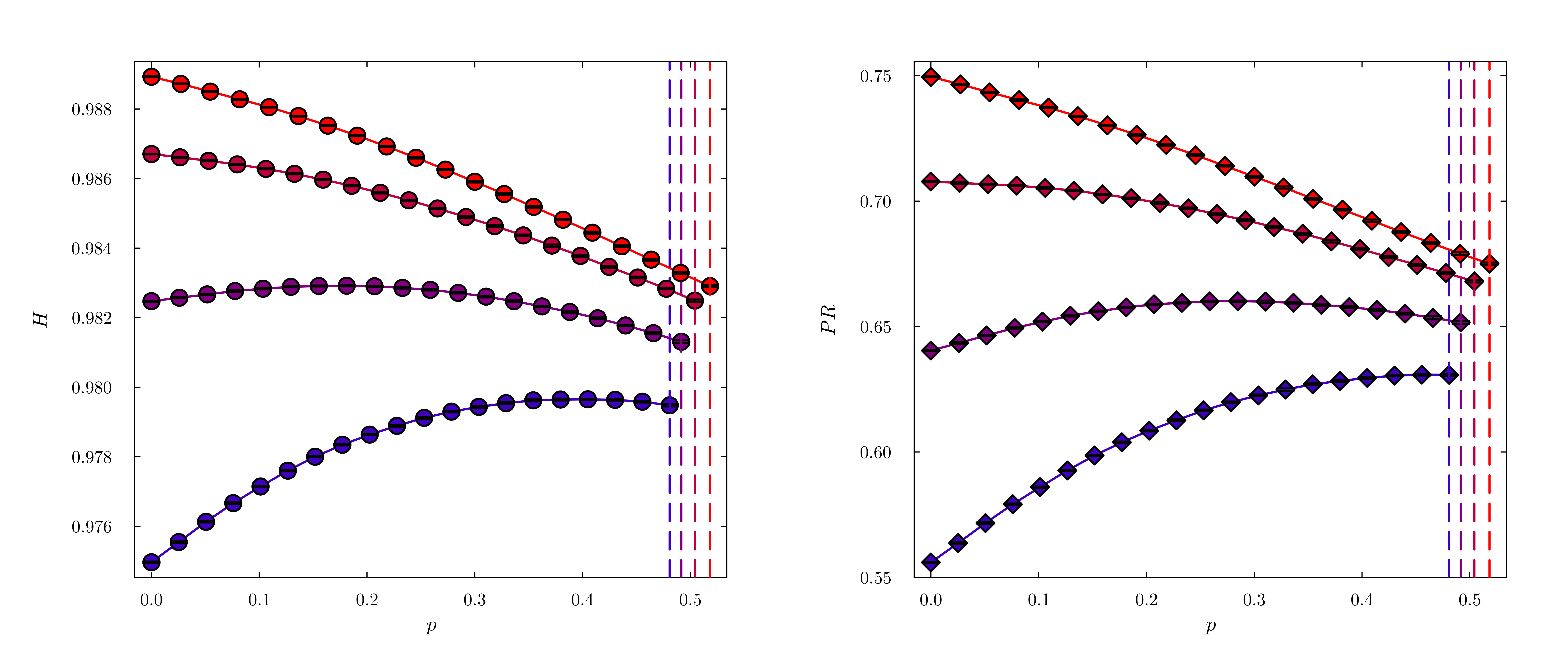}

        \caption{$\beta=0.1$.}

        \label{fig:PR_beta01}

    \end{subfigure}

    \vspace{0.5em}

    \begin{subfigure}{\textwidth}

        \centering

        \includegraphics[width=\linewidth]{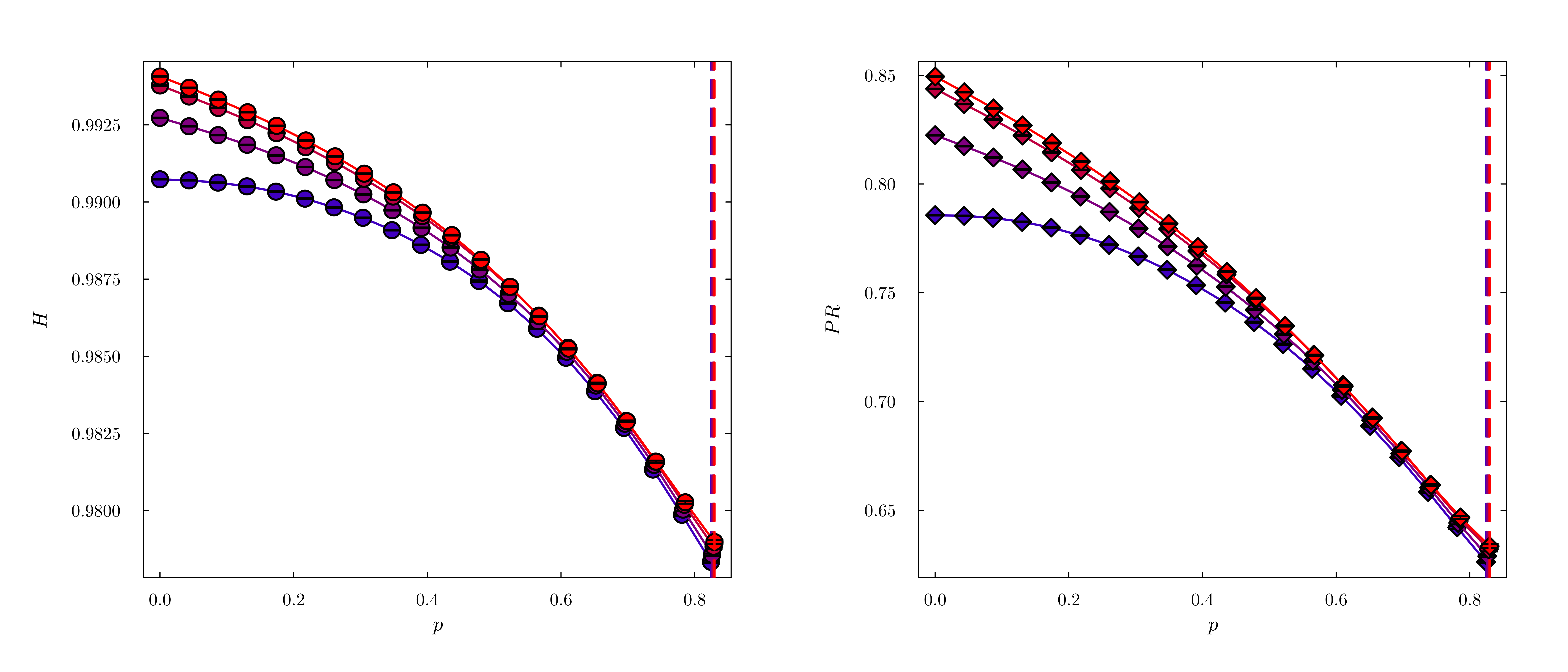}

        \caption{$\beta=1.0$.}

        \label{fig:PR_beta10}

    \end{subfigure}

    \caption{
    Comparison of entropy and the normalised participation ratio (PR) as control parameters for traffic delocalisation in the Newman--Watts model with different shortcut probabilities.
    }
    \label{fig:PR}

\end{figure}